\documentclass[reprint, amsmath,amssymb,aps,prb,superscriptaddress]{revtex4-2}   	
\usepackage{graphicx}	
\usepackage{dcolumn}
\usepackage{subfigure}
\usepackage{titlesec}
\usepackage{mathtools}
\usepackage{physics}
\usepackage{bm}
\usepackage{tikz}
\usepackage{array}
\usepackage{anyfontsize}
\usepackage{t1enc}
\usepackage{color,soul}
\usepackage{tabu}
\usepackage{lipsum}
\usepackage{hyperref}
\usepackage{longtable}

\newcommand{\rom}[1]{\textup{\uppercase\expandafter{\romannumeral#1}}}

\newcommand{\ba}{\begin{array}}
\newcommand{\ea}{\end{array}}
\newcommand{\be}{\begin{equation}}
\newcommand{\ee}{\end{equation}}
\newcommand{\bd}{\begin{displaymath}}
\newcommand{\ed}{\end{displaymath}}
\newcommand{\bi}{\begin{itemize}}
\newcommand{\ei}{\end{itemize}}
\newcommand{\bn}{\begin{enumerate}}
\newcommand{\en}{\end{enumerate}}
\newcommand{\pa}{\partial}
\newcommand{\f}{\frac}
\newcommand{\mb}{\mathbf}

\begin{document}

\title{Spin-injection-generated shock waves and solitons in a
  ferromagnetic thin film: the spin piston problem}

\author{Mingyu Hu}
\email{mingyu.hu@colorado.edu}
\affiliation{Department of Applied Mathematics, University of Colorado, Boulder, CO 80309, USA}

\author{Ezio Iacocca}
\affiliation{Department of Physics \& Energy Science, University of Colorado, Colorado Springs, CO 80918, USA}

\author{Mark Hoefer}
\affiliation{Department of Applied Mathematics, University of Colorado, Boulder, CO 80309, USA}

\date{\today}

\begin{abstract}
  The unsteady, nonlinear magnetization dynamics induced by spin
  injection in an easy-plane ferromagnetic channel subject to an
  external magnetic field are studied analytically.  Leveraging a
  dispersive hydrodynamic description, the Landau-Lifshitz equation is
  recast in terms of hydrodynamic-like variables for the
  magnetization's perpendicular component (spin density) and azimuthal
  phase gradient (fluid velocity). Spin injection acts as a moving
  piston that generates nonlinear, dynamical spin textures in the
  ferromagnetic channel with downstream quiescent spin density set by
  the external field.  In contrast to the classical problem of a
  piston accelerating a compressible gas, here, variable spin
  injection and field lead to a rich variety of nonlinear wave
  phenomena from oscillatory spin shocks to solitons and rarefaction
  waves. A full classification of solutions is provided using
  nonlinear wave modulation theory by identifying two key aspects of
  the fluid-like dynamics: subsonic/supersonic conditions and
  convex/nonconvex hydrodynamic flux.  Familiar waveforms from the
  classical piston problem such as rarefaction (expansion) waves and
  shocks manifest in their spin-based counterparts as smooth and
  highly oscillatory transitions, respectively, between two distinct
  magnetic states.  The spin shock is an example of a dispersive shock
  wave, which arises in many physical systems.  New features without a
  gas dynamics counterpart include composite wave complexes with
  "contact'' spin shocks and rarefactions.  Magnetic supersonic
  conditions lead to two pronounced piston edge behaviors including a
  stationary soliton and an oscillatory wavetrain. These coherent wave
  structures have physical implications for the generation of high
  frequency spin waves from pulsed injection and persistent, stable
  stationary and/or propagating solitons in the presence of magnetic
  damping.  The analytical results are favorably compared with
  numerical simulations.
\end{abstract}

\maketitle

\section{Introduction}

Spin transport in magnetic materials has been intensely studied, due,
in part, to its potential spintronic applications in information
technology.
A promising means for long-distance transport of angular momentum is
by way of large-amplitude, fluid-like excitations
\cite{iacocca2019,sonin_superfluid_2020}.  A useful approach to study
these nonlinear spin dynamics is the hydrodynamic framework. First
proposed by Halperin and Hohenberg \cite{halperin1969} to describe
spin waves in anisotropic ferro- and antiferromagnets under
long-wavelength assumptions, the hydrodynamic
perspective---essentially a transformation of the Landau-Lifshitz
equation to a set of fluid-like variables---has since been utilized by
a number of researchers to investigate a variety of novel spin
textures and dynamics, sometimes referred to as superfluid spin
transport
\cite{konig2001,sonin2010,takei2014,chen2014,iacocca2017,iacocca2017symmetry,iacocca2019def,evers2020,hu2021,smith2021}.
Actually, magnetic damping implies energy dissipation, which must be
compensated if sustained superfluid-like spin states are desired.  An
adequate compensation mechanism is the injection of spin into material
boundaries via the spin-Hall effect, spin-transfer torque
\cite{takei2014,chen2014,iacocca2017,stepanov2018,schneider2018,iacocca2019def},
or the quantum spin-Hall effect \cite{yuan2018}.  Recent experimental
observations of superfluid-like spin transport indicate that such
dynamics are possible \cite{stepanov2018,yuan2018}.

\begin{figure*}
  \centering
  \includegraphics[scale=0.45]{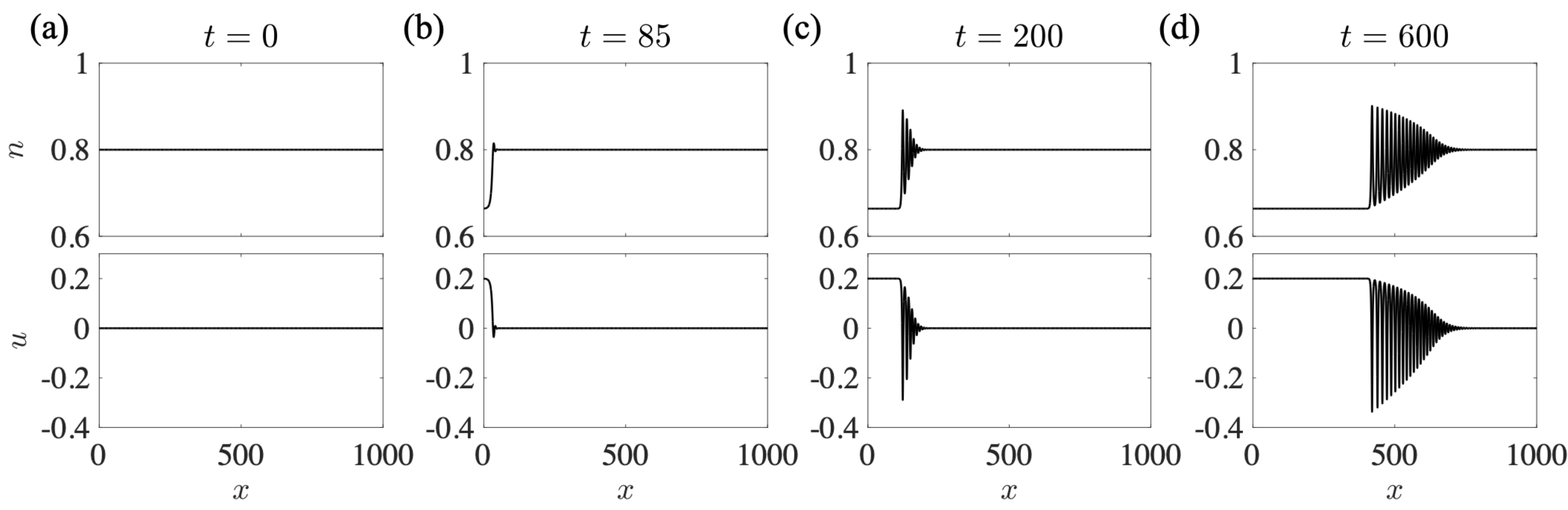}
  \caption{Temporal development of a spin shock (DSW) solution to the
    piston problem in a 1D easy-plane ferromagnetic channel. (a)
    Initial state prior to spin piston acceleration with constant spin
    density $n = h_0 = 0.8$ and fluid velocity $u = 0$.  (b) The
    piston velocity accelerates to $99.9\%$ of $u_0 = 0.2$ leading to
    a compressive wave. (c) A spin shock is under development. (d) A spin shock is fully developed. }
  \label{fig:dsw_development}
\end{figure*}
The analytical study of fluid-like spin transport in ferromagnetic
materials can be conveniently formulated in terms of dispersive
hydrodynamics (DH) in which large scale, nonlinear wave motion in a
dispersive medium is described by conservation laws subject to
dispersive corrections \cite{el2016}.  Just such a DH formulation of
magnetization dynamics was proposed in \cite{iacocca2017} as an exact
transformation of the Landau-Lifshitz (LL), a standard continuum,
micromagnetic model of ferromagnetic materials. It recasts the
three-component magnetization vector $\mathbf{m} = (m_x,m_y,m_z)$,
constrained to normalized unit length, in terms of two dependent
variables: the longitudinal spin density $n = m_z$ and the fluid
velocity $\mathbf{u} = - \nabla \arctan(m_y/m_x)$.  The latter is
proportional to the longitudinal component of the spin current.  Since
$m_y,m_x \to 0$ when $m_z \to \pm 1$, $\mathbf{u}$ is undefined when
the magnetization is saturated in the perpendicular direction so this
is referred to as the vacuum state.  When the local fluid speed
exceeds the critical value $u_{\rm cr} = \sqrt{(1-n^2)/(1+3n^2)}$
(different from the local speed of sound due to broken Galilean
invariance), the flow can be understood as supersonic
\cite{iacocca2017}, resulting, for example, in the generation of
magnetic vortices and antivortices with vacuum states at their core
\cite{iacocca_vortex-antivortex_2017,iacocca_controllable_2020}. Since
the transformation is exact, the DH formulation captures all of the
essential physics that are involved: exchange, anisotropy, and
damping, manifesting as dispersion, nonlinearity, and viscous effects,
respectively. Under conditions in which anisotropy and exchange
dominate, such as when the spin density exhibits a large gradient, the
spin system can develop rapidly oscillatory structures including
solitons and spin shocks, also known as dispersive shock waves (DSWs),
observed in the envelope of magnetostatic spin waves in Yttrium Iron
Garnet films \cite{janantha2017}.  Such DSWs are known to occur in a
variety of other DH media including Bose-Einstein condensates (BECs)
\cite{hoefer_dispersive_2006-1}, nonlinear spatial
\cite{bienaime_quantitative_2021} and fiber \cite{xu_dispersive_2017}
optics, and fluid dynamics
\cite{trillo_observation_2016,maiden_observation_2016}.  Highly
oscillatory, unsteady DSWs contrast sharply with nonlinear,
dissipation-dominated, non-oscillatory, steady shock waves in other
systems such as a compressible gas \cite{liepmann_elements_1957}.

With the DH interpretation of spin dynamics in mind, we will focus on
the analytical study of the canonical problem of spin injection into
an easy-plane ferromagnet as a feasible mechanism to generate
large-amplitude, unsteady spin textures with fluid-like features.
This problem was recently considered by us in \cite{hu2021} by way of
numerical simulations of the LL equation.  We showed that the presence
of a perpendicular, uniform, external magnetic field and the rapid
onset of spin injection resulted in three evolutionary stages: 1)
\textit{injection rise} leading to the generation of fluid-like
expansion and/or compression waves, 2) \textit{pre-relaxation} in
which the dynamics are dominated by exchange and anisotropy resulting
in rarefaction and shock waves, and 3) \textit{relaxation} to steady
state where damping, exchange and anisotropy result in steady,
precessional dissipative exchange flows \cite{iacocca2019def}, also
known as spin superfluids \cite{sonin2010,takei2014}.
 
In the present work, we focus on the pre-relaxation stage where
magnetic damping is negligible relative to exchange and anisotropy.
We interpret spin injection at one material boundary as a ``spin
piston'' whose resultant spin current is analogous to the piston
velocity.  The rapid onset of spin injection causes the acceleration
of the spin piston, which in turn leads to the development of a large
spin density gradient in the sample. Such conditions result in
dispersive hydrodynamics. The piston drives fluid-like excitations
into an otherwise static magnetic configuration whose spin density is
determined by a perpendicular external magnetic field.  Different spin
injection and field strengths lead to a variety of spin rarefaction
waves, spin shocks, and solitons.  

\begin{figure*}
  \centering
  \includegraphics[scale=0.35]{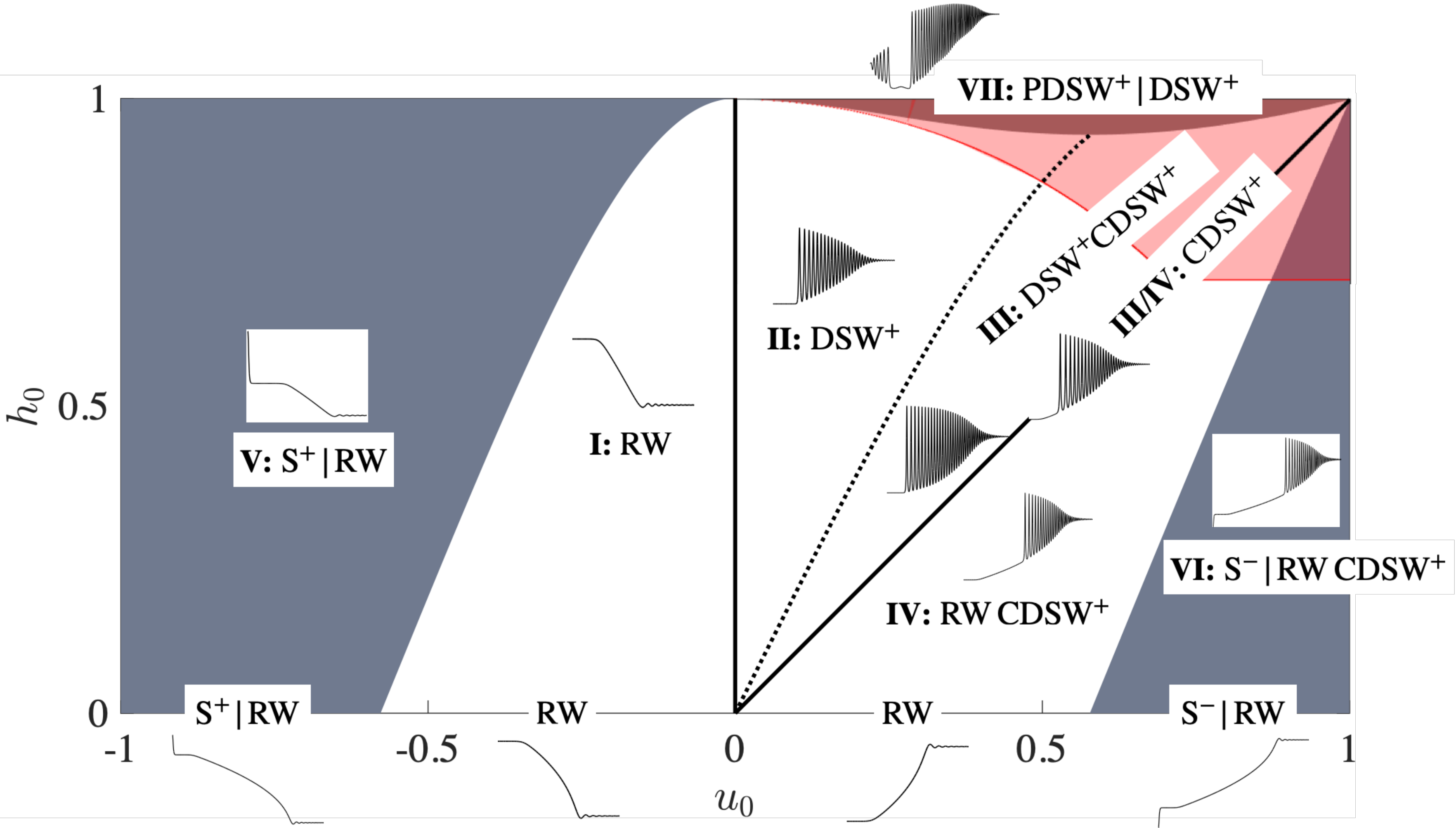}
  \caption{Classification of spin piston dynamics in terms of the
    piston velocity $u_0$ (spin injection) and background spin density
    $h_0$ (external magnetic field).  The acronyms used in the Figure
    and throughout the text are defined in Table \ref{table:acronyms}.  Dotted
    black curve: divide between convex (left) and nonconvex (right)
    regimes. Solid black lines: $r_+^L = r_+^R$ (see main text). The
    pink-shaded region implies the existence of the vacuum state $|n|
    = 1$ within the oscillatory solution. The subsonic regime is
    identified in white. Sector $\rom{1}$: $\mathrm{RW}$; Sector
    $\rom{2}$: $\mathrm{DSW}^+$; Sector $\rom{3}$: $\mathrm{DSW}^+
    \mathrm{CDSW}^+$; Boundary $\rom{3}/\rom{4}$ between sector
    $\rom{3}$ and $\rom{4}$: $\mathrm{CDSW}$; Sector $\rom{4}$:
    $\mathrm{RW}\,\mathrm{CDSW}^+$. The supersonic regime is in the
    gray, shaded region. Sector $\rom{5}$: $\mathrm{S}^+|\mathrm{RW}$,
    supersonic condition $v_- < v_+ < 0$; Sector $\rom{6}$:
    $\mathrm{S}^-|\mathrm{RW}\,\mathrm{CDSW}^+$, supersonic condition
    $v_- < v_+ < 0$; Sector $\rom{7}$:
    $\mathrm{PDSW}^+|\mathrm{DSW}^+$,
    supersonic condition $0 < v_- < v_+$.}
  \label{fig:phase_diagram}
\end{figure*}

The development of a spin shock generated by the spin piston is shown
in Fig.~\ref{fig:dsw_development}.  As demonstrated in \cite{hu2021},
by considering the problem on short enough time scales, we can neglect
magnetic damping.  Therefore, in this work, we use nonlinear wave/Whitham modulation
theory \cite{whitham2011,kamchatnov2000,el2016} to analytically
classify the dynamic spin textures generated by the spin piston with
fixed velocity $u_0$ and field $h_0$ with negligible damping.  Our primary result is the phase
diagram depicting the various solution types in the injection-field ($u_0$-$h_0$) plane
of Fig.~\ref{fig:phase_diagram}.  This diagram demonstrates the rich
selection of fluid-like spin textures that can be generated in this
system.  Moreover, these dispersive hydrodynamic waves have physical
implications for the generation of spin waves from pulsed injection
and stable solitons coincident with dissipative exchange flows.


In addition to the piston problem, another canonical hydrodynamic problem is
the space-time evolution of an initial, sharp gradient, known as the
Riemann problem \cite{riemann_uber_1860}.  We highlight the work of
\cite{ivanov2017dsw} in which the Riemann problem for polarization
waves in a two-component BEC is classified.  As it turns out, the
governing equations studied there are equivalent to the LL equation in
one spatial dimension that we study here in dispersive hydrodynamic
form absent magnetic damping.  As such, we rely heavily upon the
analysis carried out in \cite{ivanov2017dsw}.  Nevertheless, the
piston problem studied here introduces new boundary effects that do
not occur in Riemann problems such as supersonic flow conditions that
generate a soliton attached to the piston or a partial DSW that
emanates from it.  Moreover, the spin piston problem is a physically
plausible setting to generate spin shocks and other dispersive
hydrodynamic spin textures in magnetic materials.

\begin{table}
\caption{List of acronyms and symbols.}
\begin{ruledtabular}
\begin{tabular}{ l|l } 
 RW & rarefaction wave  \\ 
 \hline
 DSW & dispersive shock wave  \\ 
  \hline
 CDSW & contact dispersive shock wave  \\ 
  \hline
 PDSW & partial dispersive shock wave \\
  \hline
 S & soliton \\
  \hline
 |  & constant plateau separating waves \\
  \hline
  $\pm$ superscripts & $+$: elevation soliton, $-$: depression soliton
 \end{tabular}
 \end{ruledtabular}
 \label{table:acronyms}
\end{table}
Related superfluid and superfluid-like piston problems have been
studied theoretically \cite{hoefer_piston_2008,kamchatnov_flow_2010}
and experimentally \cite{mossman_dissipative_2018,bendahmane2020} in
BECs and optics.  While they reveal intriguing dispersive hydrodynamic
features such as the generation of an oscillatory wake at the piston
accompanied by vacuum points \cite{hoefer_piston_2008,bendahmane2020},
there are a number of new effects predicted by the spin piston problem
studied here.  This is because the hydrodynamic flux of the spin
system is \textit{nonconvex} whereas the flux in BEC and optics is
convex.  Nonconvexity manifests in the spin analogues of conservation
of mass and momentum with non-monotonic hydrodynamic fluxes are in
the spin density and fluid velocity.  Mathematically, the
long-wavelength hydrodynamic system loses strict hyperbolicity and/or
genuine nonlinearity.  This leads to new types of dispersive
hydrodynamics \cite{el2017nonconvex}.
In addition to expanding rarefaction waves (RWs) and compressive DSWs
in Figures \ref{fig:dsw_development} and \ref{fig:phase_diagram},
nonconvexity results in hybrid spin textures composed of a RW and a
special kind of contact spin shock or contact DSW (CDSW)---the
dispersive hydrodynamic analogue of a contact discontinuity in gas
dynamics---whose velocity coincides with a long wavelength magnetic
speed of sound.  Table \ref{table:acronyms} lists the acronyms and symbols used
throughout the main text and in Fig.~\ref{fig:phase_diagram}.  We
identify the supersonic transition at the piston as coincident with
either the rapid generation of a stationary soliton or a partial DSW.
Finally, sufficiently large external field and positive piston
velocity result in the generation of vacuum points within the
oscillatory solution.  Although we focus on the early, dissipationless
spin dynamics, each of these distinct dispersive hydrodynamic
excitations have implications for the long-time, steady-state
evolution of the spin system subject to magnetic damping
\cite{hu2021}. These implications are discussed in our concluding remarks Section \ref{sec:7}.

\begin{table}
\caption{Physical and mathematical properties of the solution sectors
  in the phase diagram of Fig.~\ref{fig:phase_diagram}.}
\begin{ruledtabular}
\begin{tabular}{ l|l } 
  I: RW & 
  \begin{tabular}{l}
  subsonic, expansive, convex
  \end{tabular} \\ 
 \hline
 II: DSW$^+$ & 
 \begin{tabular}{l}
 subsonic, compressive, convex
 \end{tabular}  \\ 
 \hline
 III: DSW$^+$ CDSW$^+$ &
 \begin{tabular}{l}
  subsonic, compressive, nonconvex
  \end{tabular}  \\ 
 \hline
 IV: RW CDSW$^+$ &
 \begin{tabular}{l}
   subsonic, expansive/compressive, \\
   nonconvex
 \end{tabular}
 \\
 \hline
 V: S$^+$|RW & 
 \begin{tabular}{l}
 supersonic, expansive, convex
 \end{tabular}  \\
 \hline
 VI: S$^-$|RW CDSW$^+$ &  
 \begin{tabular}{l}
   supersonic, expansive/compressive, \\
   nonconvex
 \end{tabular}
 \\
 \hline
 VII: PDSW$^+$|DSW$^+$ & 
 \begin{tabular}{l}
   supersonic, compressive, \\ 
   convex/nonconvex
 \end{tabular}
 \end{tabular}
 \end{ruledtabular}
 \label{table:physics}
\end{table}
The rest of the paper is organized as follows.  Section \ref{sec:2}
describes the spin piston problem setup. Section \ref{sec:3} provides
a summary of the results of Whitham modulation theory from
\cite{ivanov2017dsw} so that the analysis is self-contained. Some
additional analytical details are provided in the Appendix. In Section
\ref{sec:4}, solutions with zero applied field are presented and
analyzed in both the subsonic and supersonic regimes. In Sections
\ref{sec:5} and \ref{sec:6}, solutions arising in the presence of a
uniform perpendicular applied field are analyzed in the subsonic and
supersonic regimes, respectively. Finally, we present the conclusion
in Section \ref{sec:7}.

\section{Model}\label{sec:2}

Consider a one-dimensional, easy-plane ferromagnetic channel oriented
in the $\hat{\mb{x}}$ direction with length $L$. Spin injection is
applied to the left edge where $x=0$. The right edge at $x=L$
corresponds to a free spin boundary. The governing equation is
the non-dimensional, dissipationless LL equation, given by
\be \label{eq:LL} \pa_t \mb{m} = - \mb{m} \cross
\mb{h}_{\mathrm{eff}}, \quad x \in (0,L), t>0, \ee where \be
\mb{h}_{\mathrm{eff}} = \pa_{xx} \mb{m} - m_z \hat{\mb{z}} + h_0
\hat{\mb{z}}.\label{eq:heff} \ee Here,
$\mb{m} = \mb{M}/M_s = (m_x,m_y,m_z)$ is the normalized magnetization
vector, and $M_s$ is the saturation magnetization. The effective field
\eqref{eq:heff} is also normalized by $M_s$ and consists of exchange,
easy-plane anisotropy, and a uniform externally applied magnetic field
with constant magnitude $h_0$ along the perpendicular-to-plane
($\hat{\mb{z}}$) direction. The non-dimensionalization leading to
Eq.~(\ref{eq:LL}) is achieved by scaling time by $|\gamma| \mu_0 M_s$
and space by $\lambda_{\mathrm{ex}}^{-1}$, where $\gamma$ is the
gyromagnetic ratio, $\mu_0$ is the vacuum permeability, and
$\lambda_{\mathrm{ex}}$ is the exchange length. The dissipationless LL
serves as a valid model here considering the timescale within which
damping is not a key factor in the development of the dynamical
structures \cite{hu2021}. 
We will discuss the role of damping on longer time scales in the conclusion.

The following analysis is based on the DH formulation of
Eq.~\eqref{eq:LL} in terms of the hydrodynamic variables
\begin{align*}
  \text{spin density: } n &= m_z,\\
  \text{fluid velocity: } u &= -\pa_x \Phi = -\pa_x \arctan(m_y/m_x),
\end{align*}
where $\Phi$ is the azimuthal phase angle. The DH formulation is
given by \cite{iacocca_vortex-antivortex_2017}
\begin{subequations}
  \label{eq:LL_hydro}
  \begin{align}
    \pa_t n &= \pa_x\left[(1-n^2)u\right],\label{eq:LL_hydro_1}\\
    \pa_t u &=  \pa_x \left[ (1-u^2) n \right] - \pa_x \left (\frac{\pa_{xx}
      n}{1-n^2} - \frac{n (\pa_x n)^2}{(1-n^2)^2} \right ),
    \label{eq:LL_hydro_2}
    \\
    \label{eq:16}
    \pa_t \Phi &= h_0 - (1-u^2)n + \frac{1}{\sqrt{1-n^2}} \pa_x \left (
      \frac{\pa_x n}{\sqrt{1-n^2}} \right ),
  \end{align}
\end{subequations}
where \eqref{eq:LL_hydro_2} follows from the negative gradient of
\eqref{eq:16}.  These equations result from an exact transformation of
the LL equation \eqref{eq:LL}. Equation \eqref{eq:LL_hydro} is
analogous to the mass, momentum, and Bernoulli equations for an
inviscid, irrotational, compressible fluid. Owing to a phase
singularity, the vacuum state occurs when $|n|=1$.  Equation
\eqref{eq:LL_hydro} is invariant to the reflection transformation
\begin{equation}
  \label{eq:17}
  h_0 \to -h_0, \quad n \to -n, \quad \Phi \to -\Phi, \quad u \to -u .
\end{equation}

The boundary conditions (BCs) for Eq.~\eqref{eq:LL_hydro} are
\begin{subequations}
  \label{eq:bc}
  \begin{align}
    \pa_x n(0,t) = 0, \quad &\pa_x n(L,t) = 0, \label{eq:bc_n}\\
    u(0,t) = u_b(t), \quad &u(L,t) = 0,\label{eq:bc_u}
  \end{align}
\end{subequations}
where $u_b(t)$ models the time dependence of a perfect spin injection
source that increases from 0 to the maximum intensity $|u_0|$
monotonically and smoothly with the rise time $t_0$. We adopt a
hyperbolic tangent profile to model the injection rise:
$u(0,t) = (u_0/2) \{\tanh [(t-t_0/2)/(t_0/10)]+1\}$, where $t_0 = 80$ is
the time that the injection magnitude reaches $99.99\%$ of
$|u_0|$. 
For a typical Permalloy, this hyperbolic tangent profile produces a relatively sharp change in
the hydrodynamic variables---about $2$ ns---when compared to the
typical precessional period of spin-injected DEFs, on the order of
10--20 ns \cite{iacocca2017symmetry}. The modulationally stable
region, consisting of velocities $u$ in the interval $[-1,1]$,
corresponds to stable fluid-like configurations, so we restrict
$|u_0|<1$ \cite{iacocca2017}. The initial condition (IC) is given by
\begin{subequations}
  \label{eq:ic}
  \begin{align} 
    n(x,t=0) &= h_0, \label{eq:ic_n}\\
    u(x,t=0) &= 0, \label{eq:ic_u}
  \end{align}
\end{subequations}
with $|h_0| < 1$.  Thus, the spin injection problem can be reduced to
a piston problem: a piston at $x=0$, initially with velocity $u=0$,
is accelerated to $u = u_0$, generating a flow to the right ($u_0>0$) or left ($u_0<0$) into the
quiescent fluid with density $n=h_0$. In the rest of this work, we
will refer to this piston analogy for our interpretation of the spin
dynamics that result from the initial-boundary value problem
\eqref{eq:LL_hydro}--\eqref{eq:ic}. 
We focus on the classification of solutions when they are fully developed
such as in Fig.~\ref{fig:dsw_development}(d).

\section{Nonlinear Wave Dynamics and Whitham Modulation
  Theory}\label{sec:3}

In this section, we provide some necessary background, primarily
following \cite{ivanov2017dsw}, on Whitham modulation theory, a
powerful tool for studying multiscale nonlinear wave dynamics
\cite{whitham2011,kamchatnov2000,el2016}. Modulation theory results in
equations that describe the slow variation of nonlinear, periodic
traveling wave solutions.

\subsection{Traveling Wave Solutions}
\label{sec:trav-wave-solut}

Consider the traveling wave solutions of Eq.~\eqref{eq:LL_hydro} in
the form $n(x,t) = n(\xi)$ and $u(x,t) = u(\xi)$ with the moving
coordinate $\xi = x-Vt$, where
\begin{equation}
  \label{eq:3}
  V^2 = \frac{1}{2} \left( 1 + \sum_{i<j}^4 n_i n_j + \prod_{i}^4 n_i +  \sqrt{\prod_{i}^4 (1 - n_i^2)}\right)
\end{equation}
is the square of the wave's phase speed and $n_i$, $i = 1,2,3,4$ are wave parameters.  The positive/negative square
root corresponds to right/left-going wave solutions, respectively. By insertion of $n(\xi)$, $u(\xi)$ into \eqref{eq:LL_hydro} and
direct integration, the traveling wave
satisfies the ordinary differential equation (ODE)
\be \label{eq:potential_eqn} \left( \frac{d n}{d \xi} \right)^2 =
-R(n), \ee where $R(n) = (n-n_1)(n-n_2)(n-n_3)(n-n_4)$ is the
potential function, a quartic polynomial with zeros at $n_i$, $i =
1,2,3,4$. The velocity field $u(\xi)$ can be obtained in terms of
$n(\xi)$ and the roots $n_i$ \cite{ivanov2017dsw} so we focus on the modulation analysis
with $n$. For real, ordered $n_1 \leq n_2 \leq n_3 \leq n_4$ in the
interval $[-1,1]$, the traveling wave solution for $n$ either
oscillates within $[n_1, n_2]$ or $[n_3, n_4]$ with wavelength given
by \be L = \f{4 K(m)}{\sqrt{(n_3 - n_1)(n_4 - n_2)}}, \label{eq:L_ni}
\ee where $K(m)$ is the complete elliptic integral of the first kind
with $m$ given by \be m = \frac{(n_4 - n_3)(n_2 - n_1)}{(n_4 -
  n_2)(n_3 - n_1)}. \label{eq:m_ni} \ee

When $n$ oscillates within $[n_1, n_2]$, the traveling wave solution
is
\begin{equation}
  \label{eq:21}
  n(\xi) = n_2 - \frac{(n_2 - n_1)
    \mathrm{cn}^2(W,m)}{1+\frac{n_2 - n_1}{n_4 -
      n_2}\mathrm{sn}^2(W,m)}, 
\end{equation}
where 
\be 
W = \sqrt{(n_3 - n_1)(n_4 - n_2)}\xi/2
\label{eq:W}
\ee
and $\mathrm{cn}$, $\mathrm{sn}$ are Jacobi elliptic functions
\cite{byrd_handbook_1954}. The velocity is given in terms of $n$ by
\be 
u = -\frac{A_1 + V n}{1 - n^2},
\label{eq:u_ni}
\ee
where
\be 
\begin{aligned}
A_1^2 &=\frac{1}{2}\left(1 + \displaystyle \sum_{i<j}^4 n_i n_j + \prod_i^4 n_i \mp \prod_i^4 \sqrt{1-n_i^2}\right),\\
V &= \sqrt{\frac{1}{2}\left(1 + \displaystyle \sum_{i<j}^4 n_i n_j + \prod_i^4 n_i \pm \prod_i^4 \sqrt{1-n_i^2}\right)}.
\label{eq:u_ni_param}
\end{aligned}
\ee
The positive square root of $A_1$ is taken here. The upper (lower) sign in \eqref{eq:u_ni_param} gives the fast (slow) branch of wave. $V$ in Eq.~\eqref{eq:3} is the fast branch. 
We can also describe the solution in terms
of an alternative set of physical wave parameters $(\bar{n},\bar{u},a,k)$, equivalent to $n_i$, corresponding to the mean spin density $\bar{n}$, mean velocity $\bar{u}$, the amplitude $a = n_2 - n_1$, and the wavenumber $k = 2 \pi /L$.

When $n_3 \rightarrow n_2$ and $m \rightarrow 1$, the solution limits
to a depression soliton 
\begin{equation}
  n = n_2 - \frac{n_2 - n_1}{\cosh ^2 W +
    \frac{n_2 - n_1}{n_4 - n_2} \sinh^2 W} \label{eq:n_exp_dark},
\end{equation}
with background mean density $\bar{n} = n_2$. The fluid velocity in the soliton limit is
\begin{equation}
  u = - \frac{B + c_s n}{1-n^2}, \label{eq:u_exp_dark}
\end{equation}
where
\begin{equation}
  \label{eq:exp_dark_param}
  \begin{split}
    B &=  \bar{u} (\bar{n}^2 - a \bar{n} +1) + \bar{n} \mu,\\
    c_s &= \bar{u} (2 \bar{n} - a) + \mu,\\
    \mu &= \pm \sqrt{(1-(\bar{n} -a)^2)(1-\bar{u}^2)},
  \end{split}
\end{equation}
where $c_s$ is the soliton speed and $\bar{u}$ is the background mean velocity. The sign $+$($-$) gives the fast (slow) soliton. In terms of the roots $\{n_i\}_{i=1}^4$, $c_s$ and $\bar{u}$ for the soliton can be obtained by taking the limit $n_3 \rightarrow n_2$ in \eqref{eq:u_ni} and \eqref{eq:u_ni_param}.


When $n_4 \to n_3$, $m \to 0$, there are two possible limiting
solutions.  If $n_2 - n_1 \ll n_3 - n_1$, then the solution limits to
a small-amplitude harmonic wave.  If $n_2 \rightarrow n_3 = n_4$, the
solution limits to a depression algebraic soliton 
\be n(x,t) = n_2 -
\frac{n_2 - n_1}{1 + \frac{1}{4} (n_2 - n_1)^2 \xi^2}. \label{eq:n_alg_dark} 
\ee 
The background mean density for
the algebraic soliton is $\bar{n} = n_2 = n_3 = n_4$. The algebraic soliton amplitude is $a = n_2 - n_1$. The background mean velocity $\bar{u}$ and algebraic soliton speed are obtained by setting $n_2 = n_3 = n_4$ in \eqref{eq:u_ni} and \eqref{eq:u_ni_param}.

When $n$ oscillates within $[n_3, n_4]$, the traveling wave solution
is
\begin{equation}
  \label{eq:22}
   n(\xi) = n_3 + \frac{(n_4 - n_3)
  \mathrm{cn}^2(W,m)}{1+\frac{n_4 - n_3}{n_3 -
    n_1}\mathrm{sn}^2(W,m)}, 
\end{equation}
where $W$ is given in \eqref{eq:W}. The velocity is \eqref{eq:u_ni} with $A_1$ taking the negative square root and the same $V$ in \eqref{eq:u_ni_param}.

When $n_3 \rightarrow n_2$ and $m \rightarrow 1$, the solution limits
to an elevation soliton 
\begin{equation}
  n = n_3 + \frac{n_4 - n_3}{\cosh ^2 W +
    \frac{n_4 - n_3}{n_3 - n_1} \sinh^2 W} \label{eq:n_exp_bright}.
\end{equation}
The background mean density for the soliton is $\bar{n}
= n_2 = n_3$. The soliton amplitude is $a =
n_4 - n_3$. The background mean velocity $\bar{u}$ is obtained by taking the limit $n_3 \rightarrow n_2$ in \eqref{eq:u_ni} and \eqref{eq:u_ni_param}. Alternatively, the fluid velocity in the soliton limit is \eqref{eq:u_exp_dark} with 
\begin{equation}
  \label{eq:exp_bright_param}
  \begin{split}
    B &=  \bar{u} (\bar{n}^2 + a \bar{n} +1) + \bar{n} \mu,\\
    c_s &= \bar{u} (2 \bar{n} + a) + \mu,\\
    \mu &= \pm \sqrt{(1-(\bar{n} + a)^2)(1-\bar{u}^2)},
  \end{split}
\end{equation}
where $+$($-$) gives the fast (slow) soliton.


When $n_2 \to n_1$, $m \to 0$, there are two possible limiting
solutions.  If $n_4 - n_3 \ll n_4 - n_1$, then the solution limits to
a small-amplitude harmonic wave.  If $n_3 \rightarrow n_2 = n_1$, the
solution limits to an elevation algebraic soliton \be n(x,t) = n_3 +
\frac{n_4 - n_3}{1 + \frac{1}{4} (n_4 - n_3)^2
  \xi^2}. \label{eq:n_alg_bright} \ee 
The background mean density for the algebraic soliton is $\bar{n} = n_3 = n_2 = n_1$ and the soliton amplitude is $a = n_4 - n_3$. The background mean velocity $\bar{u}$ and algebraic soliton speed are obtained by setting $n_1 = n_2 = n_3$ in \eqref{eq:u_ni} and \eqref{eq:u_ni_param}.

\subsection{Whitham Modulation Equations}
\label{sec:whith-modul-equat}

The modulation equations can be expressed in diagonal form by
introducing new modulation variables $\lambda_i$ known as Riemann
invariants
\begin{equation}
  \label{eq:whitham_eqn} 
  \frac{\pa \lambda_i}{\pa t} + v_i \f{\pa \lambda_i}{\pa x} = 0,\quad
  i = 1,2,3,4,
\end{equation}
where the Riemann invariants are ordered as $\lambda_1 \leq \lambda_2
\leq \lambda_3 \leq \lambda_4$, and the $v_i$ are the Whitham
velocities \be \label{eq:whitham_velocity} v_i =
\frac{1}{2}\sum_{i=1}^{4} \lambda_i - \frac{L}{2 \pa L/\pa \lambda_i},
\quad i \in {1,2,3,4}, \ee where $L = \f{4 K(m)}{\sqrt{(\lambda_3 -
    \lambda_1)(\lambda_4 - \lambda_2)}}$ and $m = \frac{(\lambda_4 -
  \lambda_3)(\lambda_2 - \lambda_1)}{(\lambda_4 - \lambda_2)(\lambda_3
  - \lambda_1)}$.  The transformation between the $n_i$ and
$\lambda_i$ is provided in the Appendix.

The LL-Whitham modulation equations \eqref{eq:whitham_eqn} are
nonconvex, namely they can lose strict hyperbolicity (two Whitham
velocities coalesce) and/or they can lose genuine nonlinearity where
$\frac{\partial v_i}{\partial \lambda_i} = 0$ in certain parameter
regimes, resulting in a nonmonotonic dependence of the Whitham
velocity on the Riemann invariant.

\subsection{Piston Sonic and Convexity Conditions}
\label{sec:pist-supers-nonc}

The modulation equations \eqref{eq:whitham_eqn} exhibit two important
limiting simplifications. By comparing \eqref{eq:L_ni} and
\eqref{eq:m_ni}, we find that the \textit{soliton limit} ($L \to
\infty$ and $m \to 1$) occurs when $\lambda_2 \to \lambda_3$ and the
spin wave \textit{harmonic limit} ($a \to 0$ and $m \to 0$) occurs
when either $\lambda_1 \to \lambda_2$ or $\lambda_3 \to \lambda_4$. In
these limits, two of the modulation equations coincide with the
long-wave, dispersionless limit of
Eqs.~\eqref{eq:LL_hydro_1}~\eqref{eq:LL_hydro_2}
\begin{subequations}
  \label{eq:2}
  \begin{align} 
    \partial_t \bar n &= \partial_x \left[(1-\bar n^2)\bar
      u\right],\label{eq:hydro_dispersionless_1}\\ 
    \partial_t \bar u &= \partial_x \left[ (1-\bar u^2) \bar n
    \right].\label{eq:hydro_dispersionless_2}
  \end{align}
\end{subequations}
The remaining two modulation equations merge and correspond to
modulations of either the soliton amplitude or the spin wave
wavenumber.  The limiting velocities determine the motion of DSW
edges.  In the soliton limit, we have
\begin{equation}
  \label{eq:10}
  s_- \equiv \lim_{\lambda_2 \to \lambda_3} v_2 = \lim_{\lambda_2 \to \lambda_3}
  v_3 = \tfrac12(\lambda_1 + 2 \lambda_3 + \lambda_4) .
\end{equation}
In one of the harmonic limits, we have
\begin{equation}
  \label{eq:12}
  s_+ \equiv \lim_{\lambda_3 \to \lambda_4} v_3 = \lim_{\lambda_3 \to \lambda_4}
  v_4 = 2 \lambda_4 + \frac{(\lambda_2 - \lambda_1)^2}{2(\lambda_1 +
    \lambda_2 - 2 \lambda_4)} .
\end{equation}
The velocities $s_- < s_+$ are the trailing and leading edges of the
DSW.  

The dispersionless equations \eqref{eq:2} describe the evolution of
the mean density $\bar n$ and mean velocity $\bar u$.  These equations
can be expressed in diagonal form
\begin{equation}
  \label{eq:6}
  \frac{ \pa r_{\pm}}{\pa t} + v_{\pm} \f{\pa r_{\pm}}{\pa x} = 0,
  \quad  r_{\pm} = \bar u \bar n \pm \sqrt{(1-\bar u^2)(1- \bar n^2)},
\end{equation}
where the dispersionless Whitham velocities $v_+ = \tfrac12(3r_++r_-)
= 2 \bar u \bar n + \sqrt{(1-\bar u^2)(1-\bar n^2)}$, $v_- =
\tfrac12(r_++3r_-) = 2 \bar u \bar n - \sqrt{(1-\bar u^2)(1-\bar
  n^2)}$ are also the long-wavelength spin wave velocities. These
velocities are used to identify the magnetic sonic condition
\cite{iacocca2017}. The piston is
subsonic if $v_- < 0 < v_+$, when
\begin{equation}
  \label{eq:1} 
  |\bar u| < u_{\rm cr}(\bar n) = \sqrt{\frac{1-\bar n^2}{1+3 \bar n^2}},
\end{equation}
and supersonic if $v_- < v_+ < 0$ ($\bar u < - u_{\rm cr}(\bar n)$) or
$0 < v_- < v_+$ ($\bar u > u_{\rm cr}(\bar n)$).  Consequently, two different boundary behaviors will arise.


The dispersionless system \eqref{eq:2} has simple
wave solutions where only one of the Riemann invariants changes:
$(+)$-waves when $r_-$ is constant and $(-)$-waves when $r_+$ is
constant.  These solutions require the hyperbolic system
of equations \eqref{eq:2} to remain genuinely nonlinear
\cite{lax_hyperbolic_1973}, which holds so long as
\begin{equation}
  \label{eq:5}
  \bar u \ne \pm \bar n, \quad |\bar u| \ne 1, \quad |\bar n| \ne 1.
\end{equation}
These are the \textit{convexity conditions}.

\section{Phase Diagram of Figure \ref{fig:phase_diagram}}
\label{sec:regi-phase-diagr}

In this section, we provide a qualitative description of the solution
types depicted in Fig.~\ref{fig:phase_diagram} as well as a
quantitative description of the boundaries between the different
sectors.  Each distinct solution type originates from the prevailing
physical and mathematical properties of the hydrodynamic equations
\eqref{eq:LL_hydro} at the piston boundary: subsonic/supersonic flow,
compression/expansion waves, and convexity. These properties determine
the various curves partitioning the phase diagram in
Fig.~\ref{fig:phase_diagram}.  The solution type acronyms and symbols
are defined in Tab.~\ref{table:acronyms}.  Note that the reflection symmetry
\eqref{eq:17} implies that the phase diagram can be reflected in $u_0$
and $h_0$ to obtain the classification for $h_0 < 0$.  A more
detailed, quantitative description of each solution type is developed
in the next three sections.

The Whitham modulation equations \eqref{eq:whitham_eqn} are a set of
hyperbolic equations that we will solve in order to determine the
structure of solutions in the phase diagram.  The oscillatory solutions
we obtain here exhibit the following fundamental feature: they
terminate when either the wave amplitude goes to zero (the harmonic
limit) or the wavelength goes to infinity (the soliton limit).  In
both cases, the dispersionless equations \eqref{eq:2} govern the mean
density and velocity.  A general property of hyperbolic equations such
as \eqref{eq:2} is that any dynamic front adjacent to a constant
region is a simple wave \cite{courant_supersonic_1948}.  Therefore, we
can determine a relationship between the constant states to the left
and right of the RW, DSW, CDSW, etc., by holding one dispersionless
Riemann invariant constant.  For the spin piston located at the left
boundary, we will excite the fastest wave, a $(+)$-wave, in which the
Riemann invariant $r_-$ in Eq.~\eqref{eq:6} is constant across the
wave
\begin{equation}
  \label{eq:7}
  \begin{split}
    (+)\text{-wave:} \quad &n^L u^L - \sqrt{(1-(u^L)^2)(1-(n^L)^2)} \\
    = \, &n^R u^R - \sqrt{(1-(u^R)^2)(1-(n^R)^2)} .
  \end{split}
\end{equation}
The superscripts $L$ and $R$ denote the constant (mean) states to the
left and right of the wave, respectively.  In order for a $(+)$-wave
to solve the spin piston problem, we also require the RW or DSW to
propagate to the right of the boundary.  Namely, we require the
leftmost edge of the wave to have positive velocity
\begin{equation}
  \label{eq:9}
  \text{admissibility:}\quad 0 <
  \begin{cases}
    v_+(r_-^L,r_+^L), & \text{RW}, \\
    s_-(r_-^L,\lambda_2=\lambda_3,r_+^L), & \text{DSW} .
  \end{cases}
\end{equation}
It turns out that all the solutions depicted in
Fig.~\ref{fig:phase_diagram} are admissible except in the supersonic
sector VII.

The right state is constant, determined by the external magnetic field
and free spin boundary condition \eqref{eq:ic_n}, \eqref{eq:ic_u}
\begin{equation}
  \label{eq:8}
  n^R = h_0, \quad u^R = 0 .
\end{equation}
The constant left state is achieved after the piston velocity has
saturated at $t \approx t_0$ ($u_b(t) \to u_0$), provided the
admissibility condition \eqref{eq:9} is satisfied.  When the left
state is subsonic, we use \eqref{eq:7}, \eqref{eq:8}, and
\eqref{eq:bc_u} to obtain the spin density on the left
\begin{equation}
  \label{eq:13}
  \text{subsonic:} \quad n^L = h_0\sqrt{1-u_0^2} - u_0\sqrt{1-h_0^2},
  \quad u^L = u_0 .
\end{equation}
The flow is subsonic so long as \eqref{eq:1} with $\bar u \to u^L$ and
$\bar n \to n^L$ is satisfied.  The transition from subsonic to
supersonic in the phase diagram Fig. \ref{fig:phase_diagram} occurs when
\begin{equation}
  \label{eq:14}
  |u^L| = u_{\rm cr}(n^L) .
\end{equation}
Using \eqref{eq:13}, there are multiple solutions of
eq.~\eqref{eq:14}. The region of parameters corresponding to subsonic
conditions at the $x = 0$ boundary is the interior of the following
four curves
\begin{equation}
  \label{eq:4}
  \begin{split}
    u_0 &= \pm \sqrt{\frac{2 + h_0^2 \pm h_0 \sqrt{h_0^2+8}}{6}}, \\
    u_0 &= \sqrt{\frac{-2 + 3 h_0^2 \pm h_0\sqrt{9h_0^2 - 8}}{2}},
  \end{split}
\end{equation}
In other words, \eqref{eq:4} are the sonic curves. The subsonic region
is reflected in Fig.~\ref{fig:phase_diagram} by the unshaded and
pink-shaded regions containing sectors I--IV.

Consequently, sectors V--VII are supersonic and we need an alternative
way to determine $n^L$ because $n^L \neq u_0$ at the piston
boundary.  Sectors $\rom{5}$ and $\rom{6}$ are associated with the supersonic condition $v_- < v_+ <0$ and the way to resolve this was first identified in \cite{iacocca2019def} where a stationary spin soliton was introduced with its extremum in density and velocity centered at the piston boundary.  Thus, only half the soliton is within the domain and it was referred to as a \textit{contact soliton}.  The soliton solutions, given by the fast branch of \eqref{eq:n_exp_dark} and \eqref{eq:n_exp_bright}, provide for a rapid transition from supersonic conditions at the piston to subsonic conditions in the soliton far-field $(\overline{n},\overline{u})$.  In
order to uniquely determine the soliton, we invoke three assumptions.
First, we identify the soliton far-field with
$(\overline{n},\overline{u}) = (n^L,u^L)$.  Then, for a ($+$)-wave, we
can use Eqs.~\eqref{eq:7} and \eqref{eq:8} to determine
\begin{equation}
  \label{eq:19}
  n^L = -u^L \sqrt{1 - h_0^2} + h_0 \sqrt{1 - \left(u^L\right)^2}.
\end{equation}
Second, by equating the extreme soliton velocity $u(\xi = 0)$ in
\eqref{eq:u_exp_dark} to the piston velocity $u_0$, we have
\begin{equation}
  \label{eq:18}
  u_0 = -\frac{B}{1 - \left( n^L \pm a \right)^2},
\end{equation}
where $B$ is given in Eqs.~\eqref{eq:exp_dark_param}, \eqref{eq:exp_bright_param}, $a > 0$ is the soliton amplitude.  Finally, the soliton is stationary so that
$c_s = 0$ in \eqref{eq:exp_dark_param} or \eqref{eq:exp_bright_param}, giving
\begin{equation}
  \label{eq:20}
  u^L(2 n^L \pm a) + \sqrt{(1-(n^L \pm a)^2)(1-(u^L)^2)} = 0 .
\end{equation}
The $+ (-)$ in \eqref{eq:18} and \eqref{eq:20} correspond to a bright (dark) soliton.
For example, in the supersonic sector V, the soliton is of elevation
type so \eqref{eq:exp_bright_param} applies and the $+$ sign is taken in \eqref{eq:18} and \eqref{eq:20}. The three conditions \eqref{eq:19},
\eqref{eq:18}, and \eqref{eq:20} uniquely determine the soliton
amplitude $a$ and its far-field $(n^L,u^L)$.

In \cite{hu2021}, it was shown that this problem gives rise to
compression or expansion waves emanating from the piston depending
upon the input parameters $(u_0,h_0)$.  This is determined by whether
or not the ($+$)-wave speed $v_+$ is increasing or decreasing from
left to right during the piston acceleration period. 
\begin{equation}
 \mathrm{compression:} \quad v_+(n^L(t),u_b(t)) > v_+(h_0,0)
\label{eq:15}
\end{equation} 
implies compression waves and expansion waves otherwise.
The pure compression region is reflected in
Fig.~\ref{fig:phase_diagram} by the solid black lines $u_0 = 0$ and
$u_0 = h_0$.  
When $0 < u_0 < h_0$, the subsonic solutions involve only
DSWs. When $0 < h_0 < u_0$, the subsonic solutions involve both RWs and DSWs.  When $u_0 < 0$ or $h_0 = 0$, the subsonic solutions are RWs.

When $u_0>0$, there is another effect at play: loss of convexity
\eqref{eq:5} when $u^L = |n^L|$.  For the subsonic regime, $u^L = u_0$ and \eqref{eq:13}
implies convexity is lost when
\begin{equation}
  \label{eq:11}
  \text{loss of convexity:} \quad h_0 = 2 u_0 \sqrt{1-u_0^2} .
\end{equation}
This is the dotted curve in Fig.~\ref{fig:phase_diagram}.  To the
right of this curve, the solutions exhibit hybrid waves involving
CDSWs, and either DSWs (when $0 < u_0 < h_0$) or RWs
(when $u_0 > h_0$).

One more feature of the solutions is depicted in
Fig.~\ref{fig:phase_diagram}: vacuum points.  When $|n| = 1$, the
velocity $u$ is undefined and corresponds to the absence of fluid or
vacuum.  We find that only oscillatory solutions such as DSWs and
CDSWs, i.e., $u_0 > 0$, can result in the generation of isolated
points at which $|n| = 1$.  The threshold for this behavior is
determined by equating the extrema of the oscillation density
\eqref{eq:21} or \eqref{eq:22} with $n = \pm 1$, namely $n_j = (-1)^j$
for some root $n_j$, $j \in \{ 1,2,3,4\}$.  A quantitative
determination of this threshold requires the solution of the Whitham
modulation equations \eqref{eq:whitham_eqn}, which we undertake in the
next several sections.  The vacuum threshold is depicted in the phase
diagram \ref{fig:phase_diagram} by a solid red curve, above which the
solutions exhibit vacuum points.

In the following sections, we solve the Whitham modulation
equations to obtain the detailed structure of the shock, rarefaction,
and soliton solutions.

%

\section{Zero Applied Field}\label{sec:4}

When $h_0 = 0$, all of the dynamics are governed by the dispersionless
limit \eqref{eq:LL_hydro} with additional treatment if the solution is supersonic. This case corresponds to the horizontal axis in the phase
diagram \ref{fig:phase_diagram}. 
The $(+)$-wave for $r_+ = r_+(\xi)$ satisfies $v_+(r_-^R,r_+) = \xi = x/\left(t - \bar{t} \right)$, where $\bar{t}$ is a constant time shift, $r_-^R =-1$, and $r_+^L < r_+(\xi) < r_+^R$.

\subsection{Subsonic Regime: RW}
\begin{figure}[h!]
\centering
\includegraphics[scale=0.33]{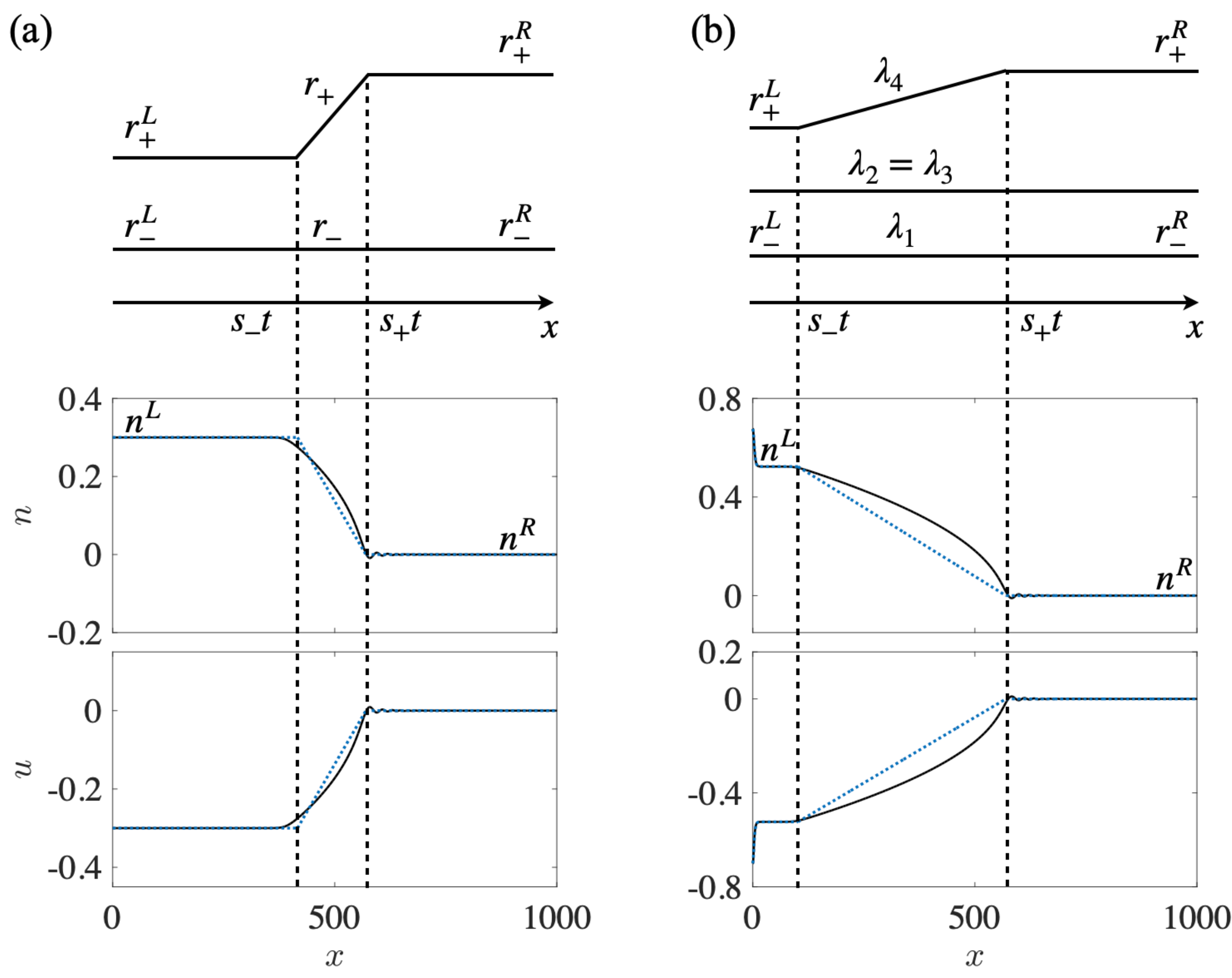}
\caption{Riemann invariant configurations in the upper panels corresponding to theoretical (dotted) and numerical (solid) solutions. The vertical dashed lines correspond to $x_{\pm} = s_{\pm}(t-\bar{t})$, where $\bar{t} = 30$ is the time delay introduced to account for the piston acceleration time. (a) RW solution with $u_0 = -0.3$, satisfying the subsonic condition $|u_0| < \frac{1}{\sqrt{3}}$; (b) soliton|RW solution with $u_0 = -0.7$, satisfying the supersonic condition $\frac{1}{\sqrt{3}} < |u_0| < 1$}
\label{fig:ri_config_rw_csrw}
\end{figure}

\begin{figure}[h!]
\centering
\includegraphics[scale=0.31]{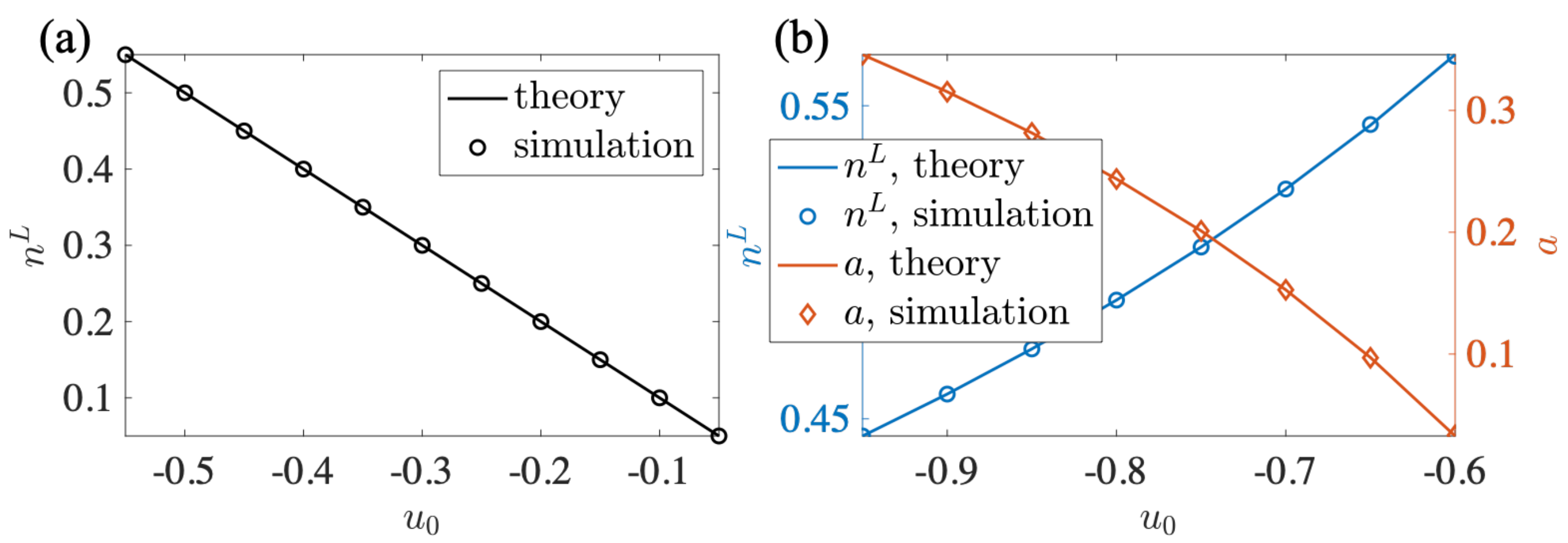}
\caption{Theory and simulation results of the left constant state density $n^L$ for subsonic $|u_0| < \frac{1}{\sqrt{3}}$ (a) and supersonic $\frac{1}{\sqrt{3}} < |u_0| < 1$ (b). The soliton amplitude $a$ is also shown in (b).}
\label{fig:err_h0_0}
\end{figure}

The subsonic solution when $h_0 = 0$ is a RW. The $(+)$-wave assumption leads to the background
spin density on the left given by \eqref{eq:13}
\be 
n^L = -u_0.  
\ee 
The system is always convex because $|u_0| < 1$ in \eqref{eq:11}.
The admissibility condition \eqref{eq:9} is satisfied until
$v_+(r_-^L,r_+^L)=0$, which is also the sonic condition \eqref{eq:14} leading to
$u_0 = \pm u_{\mathrm{cr}} = \pm \frac{1}{\sqrt{3}}$. Thus, for a RW solution to be admissible, the piston velocity $u_0$ is required to be subsonic with 
$u_0^2 < \frac{1}{3}$. We have additionally confirmed that there are no
admissible $(-)$-wave solutions with $r_+^L= r_+^R$. The Riemann invariant configuration and an example solution is shown in
Fig.~\ref{fig:ri_config_rw_csrw}(a). In the theoretical solution, a
time delay $\bar{t} = 30$ (recall, $t_0 = 80$) is introduced to account for the piston
acceleration time. This choice of time delay is consistently applied
to all the theoretical solutions in the following sections. Across the subsonic domain, our theoretical predictions on $n^L$ agree excellently with simulation results, shown in Fig.~\ref{fig:err_h0_0}(a). 


\subsection{Supersonic Regime: $\mathbf{S}^+$|RW}
When the piston velocity is supersonic with $u_0^2 > \frac{1}{3}$, a contact soliton develops at the piston boundary smoothly connected to a RW via an intermediate constant state. The Riemann invariant configuration of the solution is shown in the top panel of Fig.~\ref{fig:ri_config_rw_csrw}(b). The soliton is represented by the Riemann invariants $\lambda_2 = \lambda_3$. 

This soliton is theoretically determined by \eqref{eq:19}-\eqref{eq:20} for a bright soliton. It is verified that when the piston is moving at the sonic speed $u_0 = u_\mathrm{cr} = \pm \frac{1}{\sqrt{3}}$, the soliton does not exist, i.e. $a = 0$. Therefore, the soliton at the piston boundary only emerges in the supersonic regime. A representative supersonic solution is shown in the bottom panel of Fig.~\ref{fig:ri_config_rw_csrw}(b), exhibiting good agreement with the numerical simulation. 
Across the supersonic domain, theoretical predictions of $n^L$ and $a$ of the soliton demonstrate excellent agreement with simulation results, shown in Fig.~\ref{fig:err_h0_0}(b). Herein we have confirmed our assumptions proposed in Sec.~\ref{sec:regi-phase-diagr} on the characterization of the solitonic supersonic solutions. 
In addition, our analysis found that no vacuum state, where $|n| = 1$ occurs. The largest magnitude of $n$ is reached at the peak (crest) of the elevation (depression) soliton at the piston boundary and this magnitude is always less than 1 for $\frac{1}{3} < u_0^2 < 1$.


\begin{figure}[h!]
\includegraphics[scale=0.33]{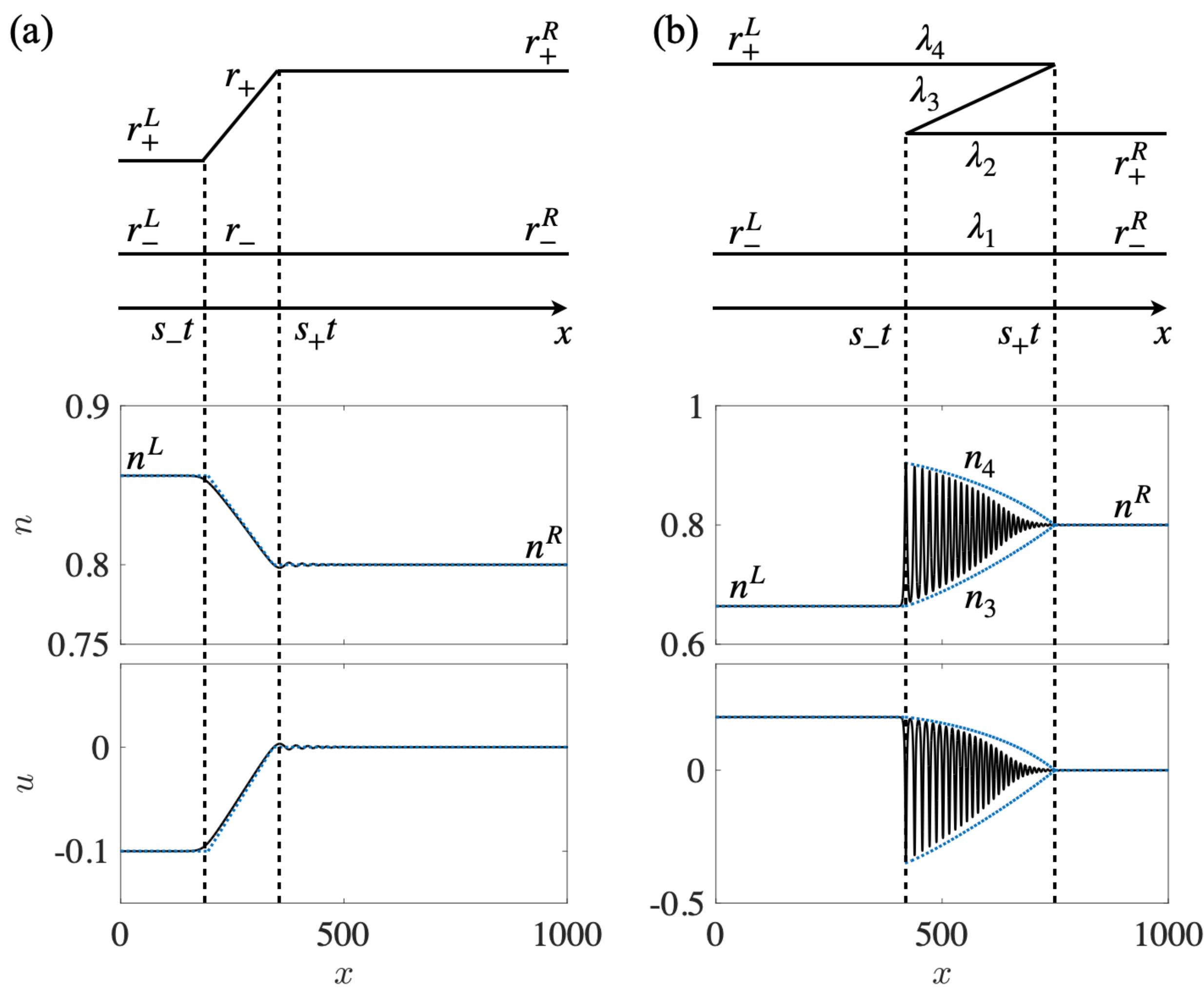}
\caption{Riemann invariant configurations and the corresponding theoretical (dotted) and numerical (solid) solutions. The vertical dashed lines correspond to $x_{\pm} = s_{\pm}(t-\bar{t})$, where $\bar{t} = 30$ is the time delay introduced to account for the piston acceleration time. (a) Sector $\rom{1}$: $\mathrm{RW}$ with $u_0 = -0.1$ and $h_0 = 0.8$. (b) Sector $\rom{2}$: $\mathrm{DSW}^+$ with $u_0 = 0.2$ and $h_0 = 0.8$, the dotted curve is the theoretical DSW envelope.}
\label{fig:convex_subsonic}
\end{figure}

\section{Nonzero Applied Field, Subsonic Regime}\label{sec:5}
In this section, we present subsonic solutions with nonzero applied field. 
The solution map is the white region of the phase diagram Fig.~\ref{fig:phase_diagram}, including sectors $\rom{1}$-$\rom{4}$. We consider each sector in turn.
\subsection{Sector $\rom{1}$: RW}
In sector $\rom{1}$, the system satisfies the convexity condition \eqref{eq:5} and yields simple wave solutions. Again, the solution is an expansive RW satisfying $r_- = r_-^R$, $r_+ = r_+(\xi)$, and $v_+(r_-, r_+) = \xi = x/ \left( t - \bar{t} \right)$. An example Riemann invariant configuration and solution in sector $\rom{1}$ is shown in Fig.~\ref{fig:convex_subsonic}(a). Good agreement between theory and simulation is demonstrated. 

The admissibility condition \eqref{eq:9} has been verified across sector $\rom{1}$. The sonic condition is determined by $v_+ = 0$, yielding the boundary between sector $\rom{1}$ and $\rom{5}$. Again, the sonic condition coincides with the admissibility threshold, indicating that only subsonic solutions are admissible in sector $\rom{1}$. On the boundary between sector $\rom{1}$ and $\rom{2}$ which is the $h_0$-axis, there are no induced dynamics because $u_0 = 0$. Furthermore, no vacuum state is present in sector $\rom{1}$ since the largest magnitude of $n$ is at the piston boundary where the piston velocity is restricted to $|u_0|<1$.


\subsection{Sector $\rom{2}$: $\mathbf{DSW}^+$}
\begin{figure}[h!]
\includegraphics[scale=0.5]{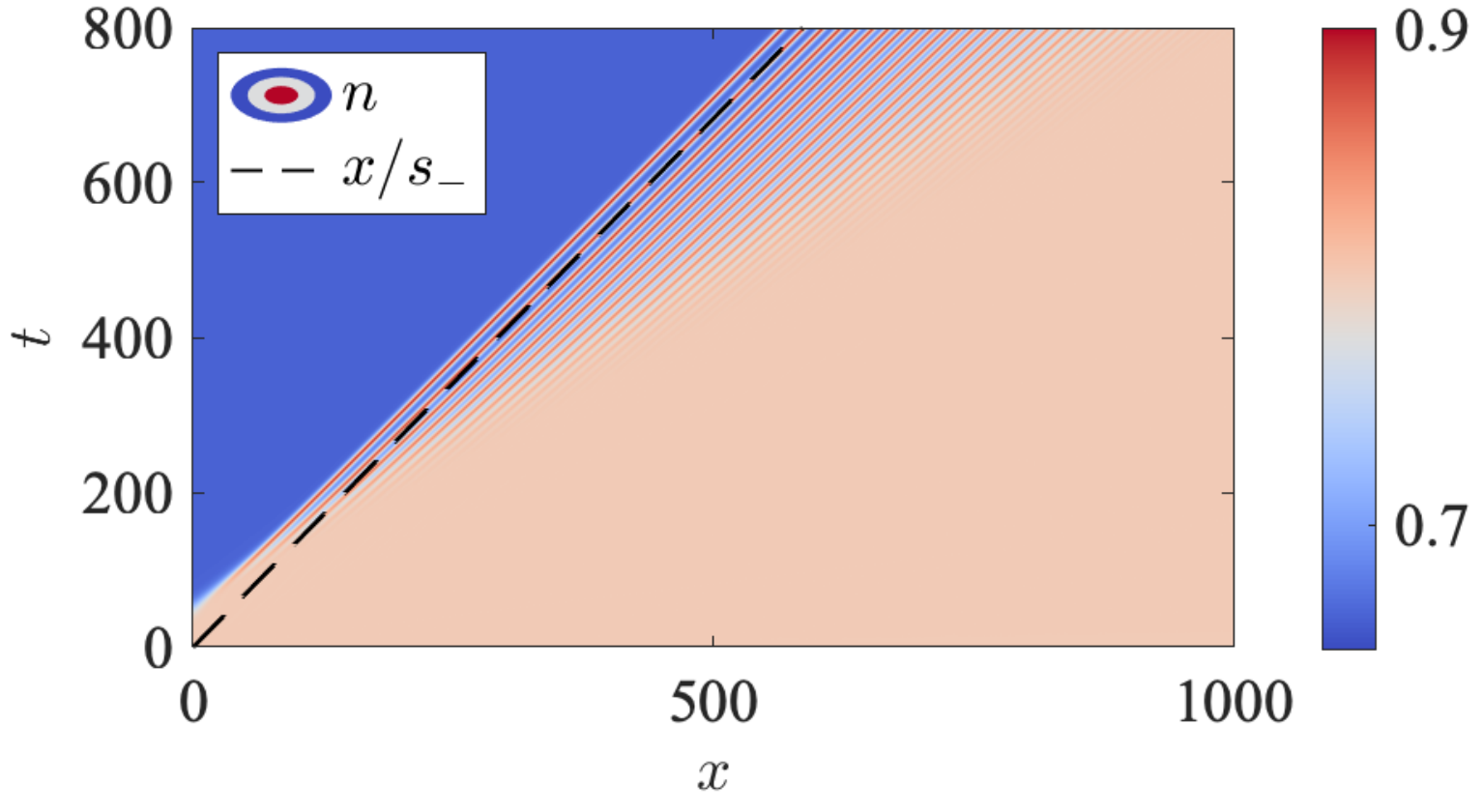}
\caption{Space-time contour plot of a $\mathrm{DSW}^+$ solution in sector $\rom{2}$ with $u_0 = 0.2$, $h_0 = 0.8$. The black solid line is the predicted trailing edge soliton location for an ideal piston with instantaneous acceleration.}
\label{fig:dsw_n_space_time_contour}
\end{figure}

Sector $\rom{2}$ is to the left of the convexity curve (dotted black curve) in the phase diagram Fig.~\ref{fig:phase_diagram}, so the system is convex, yielding simple wave solutions. Furthermore, $v_+(r_-^R,r_+^L) > v_+(r_-^R,r_+^R)$ leads to compressive DSW solutions that satisfy $\lambda_3 = \lambda_3(\xi)$, $v_3(r_-^R, r_+^R, \lambda_3,r_+^L) = \xi = x/\left(t - \bar{t} \right)$. The DSW solutions satisfy the admissibility condition ~\eqref{eq:9}. An example Riemann invariant configuration and solution in sector $\rom{2}$ is shown in Fig.~\ref{fig:convex_subsonic}(b). Near the DSW's harmonic edge, the numerical simulation and the predicted envelope amplitude deviate somewhat. This is a common feature of the asymptotic (large $t$) behavior of DSWs \cite{el2016}.  Fig.~\ref{fig:dsw_n_space_time_contour} shows that, despite the piston acceleration period, the theoretically predicted trajectory of the DSW's soliton edge differs from the simulation result by a constant time shift.

The DSW solution exhibits vacuum in the pink-shaded region in the phase diagram Fig.~\ref{fig:phase_diagram}. The vacuum state is first reached when the maximum of the trailing edge soliton density $n = n_4$ reaches 1. We evaluate $n_4$ in the soliton limit, which is a function of the Riemann invariants, to determine this threshold (see Appendix).  As time progresses, the vacuum point will move inside the oscillatory structure \cite{el2016}. Example DSWs with vacuum will be shown in  Fig.~\ref{fig:h0_nz_vac_soln_ex}. We point out that the vacuum threshold determination is the same across all subsonic sectors whose solution contains a DSW structure, despite the convexity of the system.

\subsection{Sector $\rom{3}$: $\mathbf{DSW}^+ \mathbf{CDSW}^+$}
Sector $\rom{3}$ is to the right of the convexity curve (dotted black curve) in the phase diagram Fig.~\ref{fig:phase_diagram}, so the solution breaks the convexity condition \eqref{eq:5}, manifested as the coalescence of two Riemann invariants $\lambda_3 = \lambda_4$ and Whitham velocities $v_3 = v_4$. The Riemann invariant configuration and an example solution are shown in Fig.~\ref{fig:nonconvex_subsonic}(a), satisfying $r_- = r_-^R$, $\lambda_3 = \lambda_3(\xi)$, and $v_3(r_-^R, r_+^R, \lambda_3, r_+^L) = \xi = x /\left( t - \bar{t} \right)$. The spin injection $u_0$ satisfies \eqref{eq:15}, leading to a compressive $\mathrm{DSW}^+ \mathrm{CDSW}^+$ composite wave where $r_+^L > r_+^R$ gives the DSW portion and the coalescence of Riemann invariants $\lambda_3=\lambda_4$ gives the CDSW portion.

A CDSW is a degenerate DSW solution whose soliton limit is an algebraically decaying soliton where three Riemann invariants, $\lambda_2$, $\lambda_3$, and $\lambda_4$, coincide. The algebraic soliton travels at the speed of a dispersionless (long-wave) characteristic velocity, mimicking a contact discontinuity in viscous hydrodynamics. It is observed numerically that CDSWs generally require a longer time than DSWs to develop. Therefore, a larger discrepancy between the simulated CDSW portion and the analytical wave envelope is observed compared to the DSW portion. The admissibility of the composite wave solution in sector $\rom{3}$, $0< s_-^{\mathrm{(1)}} < s_-^{\mathrm{(2)}} < s_+$, has been confirmed. The region where a vacuum state is present in the solution is shaded in pink in Fig.~\ref{fig:phase_diagram} and a typical solution is shown in Fig.~\ref{fig:h0_nz_vac_soln_ex}(c).

\subsection{Sector $\rom{4}$: $\mathbf{RW}\,\mathbf{CDSW}^+$}
\begin{center}
\begin{figure}[h!] 
\includegraphics[scale=0.33]{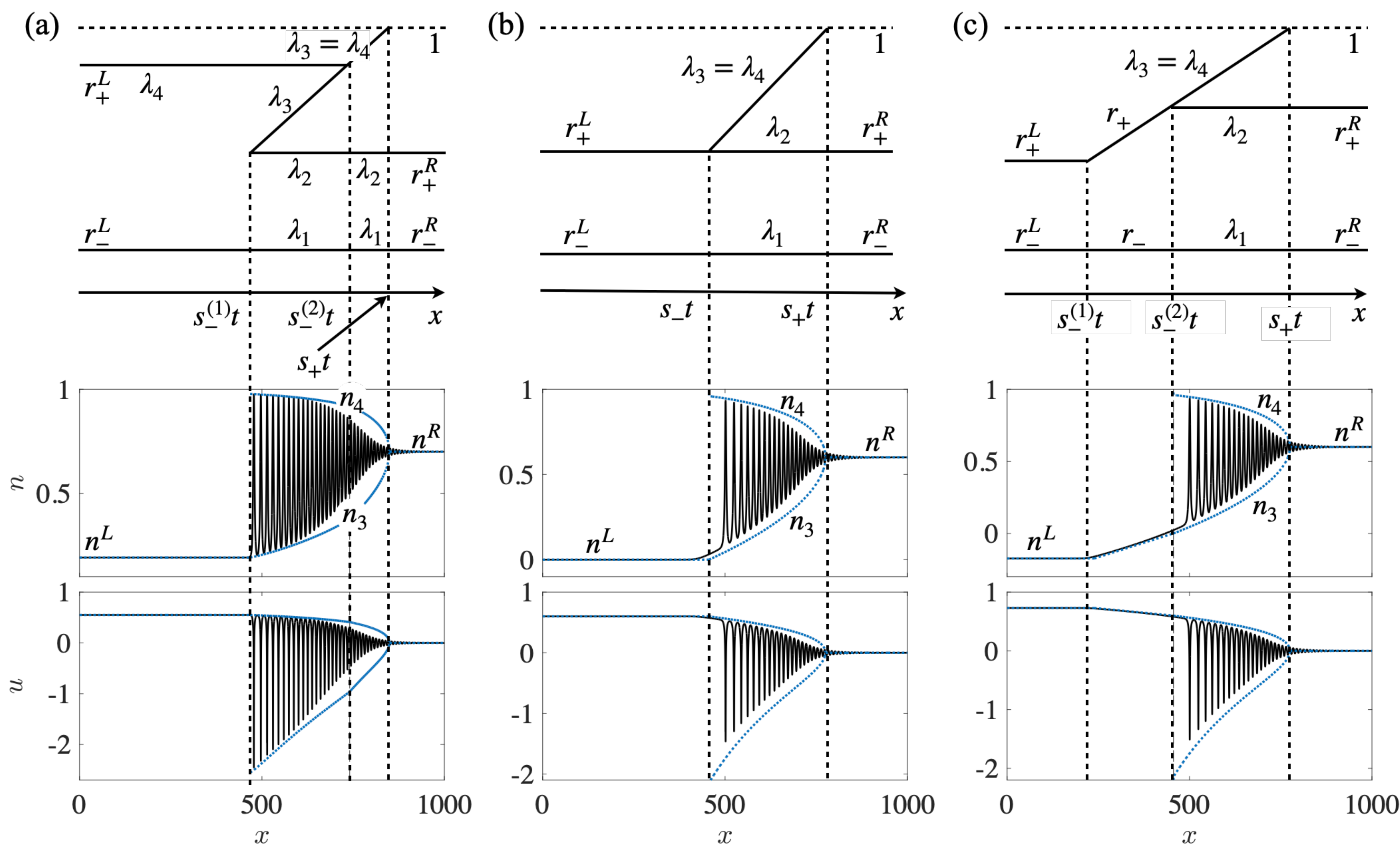}
\caption{Riemann invariant configurations and example solutions when $h_0 \neq 0$ for
(a) Sector $\rom{3}$: $\mathrm{DSW}^+ \mathrm{CDSW}^+$ with $u_0 = 0.55$ and $h_0 = 0.7$; (b) Boundary of sectors $\rom{3}$ and $\rom{4}$: $\mathrm{CDSW}^+$ with $u_0 = 0.6$ and $h_0 = 0.6$. (c) Sector $\rom{4}$: $\mathrm{RW\,CDSW}^+$ with $u_0 = 0.73$ and $h_0 = 0.6$;  The vertical dashed lines separate different components of the composite wave solutions based on predicted edge velocities. The dotted curves are the predicted envelopes of the DSW structure in the solution. In (b), the dotted curve also predicts the dispersionless RW portion of the solution. All modulation solutions include the time delay $\bar{t} = 30$ to account for the piston acceleration time.}
\label{fig:nonconvex_subsonic}
\end{figure}
\end{center}

\begin{figure}[h!]
\centering
\includegraphics[scale=0.35]{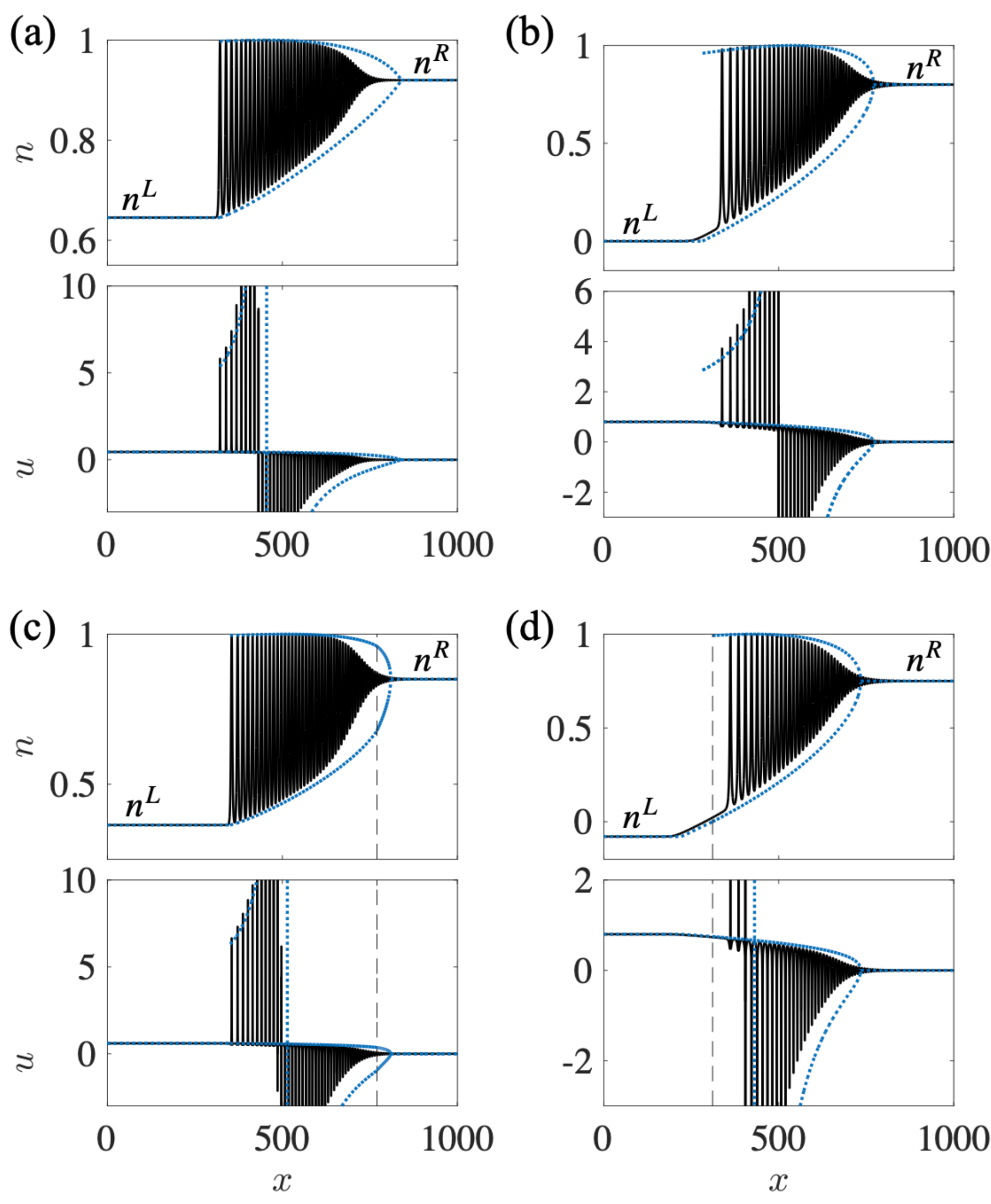}
\caption{Example solutions with vacuum states when $h_0 \neq 0$. (a) Sector $\rom{2}$: $\mathrm{DSW}^+$ with $u_0 = 0.45$ and $h_0 = 0.92$; (b) Border of sectors $\rom{2}$ and $\rom{3}$: $\mathrm{CDSW}^+$ with $u_0 = 0.8$ and $h_0 = 0.8$; (c) Sector $\rom{3}$: $\mathrm{DSW}^+ \mathrm{CDSW}^+$ with $u_0 = 0.6$ and $h_0 = 0.85$; (d) Sector $\rom{4}$: $\mathrm{RW}\,\mathrm{CDSW}^+$ with $u_0 = 0.8$ and $h_0 = 0.75$. The dotted curves are the predicted envelopes of the DSW structure. In (d), the dotted curve includes the predicted dispersionless RW portion in the solution. The vertical dashed lines separate different components in composite modulation solutions. The modulation solutions includes the time delay $\bar{t} = 30$ to account for the piston acceleration time.}
\label{fig:h0_nz_vac_soln_ex}
\end{figure}

Before moving on to sector $\rom{4}$, we discuss the solution on the boundary between sector $\rom{3}$ and $\rom{4}$, where $r_+^L = r_+^R$ as shown in the Riemann invariant configuration in Fig.~\ref{fig:nonconvex_subsonic}(b).  The system is nonconvex and the solution is a single $\mathrm{CDSW}^+$ because $\lambda_3 = \lambda_4$ across the shock. 
\onecolumngrid
\twocolumngrid
Sector $\rom{4}$ is to the right of the convexity threshold (dotted black curve, Eq.~\eqref{eq:11}) in Fig.~\ref{fig:phase_diagram}, so the system is nonconvex. During piston acceleration, compressive dynamics are induced, then followed by expansive dynamics. Thus, the solution is a $\mathrm{RW}\,\mathrm{CDSW}^+$ composite wave that satisfies $r_- = r_-^R$, $v_3 (r_-^R, r_+^R, \lambda_3,\lambda_3) = \xi = x / \left( t - \bar{t} \right)$,  and $\lambda_4 = \lambda_3$. The Riemann invariant configuration and an example solution are shown in Fig.~\ref{fig:nonconvex_subsonic}(c). The admissibility \eqref{eq:9} of the solutions have been verified in the sector with the threshold $s_-^{(1)}=0$ coinciding with the sonic condition $v_+=0$ and Eq.~\eqref{eq:14}. The vacuum region, shaded in pink in Fig.~\ref{fig:phase_diagram}, is determined by evaluating the wave envelope $n_4$ in the algebraic soliton limit. The onset of vacuum is found to be independent of $u_0$ in this case and happens at $h_0 = 1/\sqrt{2}$. Representative solutions containing a vacuum point are shown in Fig.~\ref{fig:h0_nz_vac_soln_ex}(c), (d).

In the example solutions shown in Fig.~\ref{fig:nonconvex_subsonic}(b), (c), we observe that there is a smooth tail at the algebraic soliton limit of the CDSW when it connects to the dispersionless portion of the solution. This phenomenon is most evidently shown in Fig.~\ref{fig:nonconvex_subsonic}(b) with a single CDSW. This behavior does not occur in DSWs where the exponential soliton edge terminates directly at the dispersionless edge state (see the bottom panels of Figs.~\ref{fig:convex_subsonic}(b) and \ref{fig:nonconvex_subsonic}(a)). This phenomenon serves as a distinguishing feature to identify the soliton edge of a CDSW.

\begin{figure}[h!]
\centering
\includegraphics[scale=0.31]{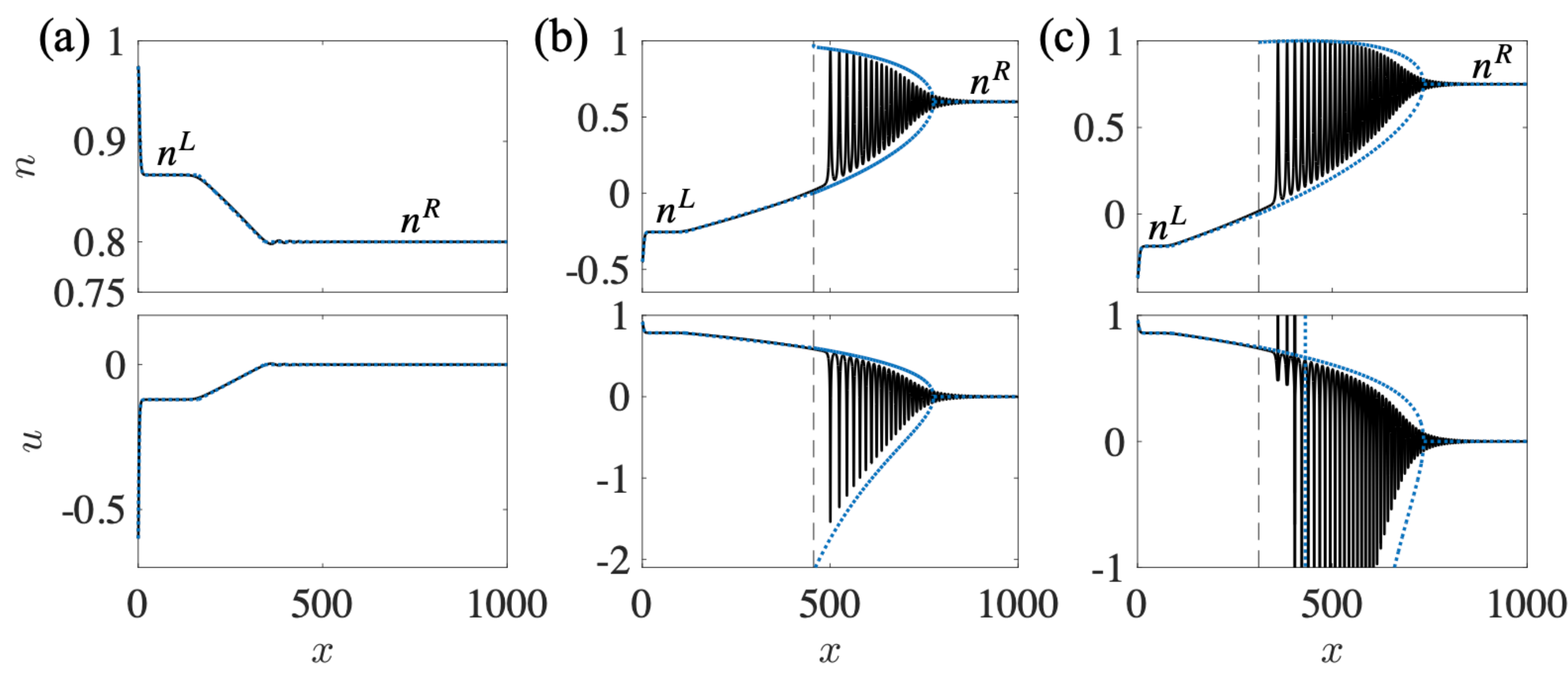}
\caption{
Supersonic solutions when $h_0 \neq 0$. (a) Sector $\rom{5}$: $\mathrm{S}^+|\mathrm{RW}$ with $u_0 = -0.6$ and $h_0 = 0.8$; (b) Sector $\rom{6}$: $\mathrm{S}^-|\mathrm{RW}\,\mathrm{CDSW}^+$ with $u_0 = 0.92$ and $h_0 = 0.6$; (c) Sector $\rom{6}$: $\mathrm{S}^-|\mathrm{RW}\,\mathrm{CDSW}^+$ with a vacuum point, $u_0 = 0.96$ and $h_0 = 0.75$. The dotted curves trace the predicted piston edge soliton, the dispersionless portion of the solution, and the envelope of the DSW-type portion of the solution. The vertical dashed lines divide the different components in a composite wave based on the predicted edge velocities. The time delay $\bar{t} = 30$ is used in theoretical plotting to account for the piston acceleration time.}
\label{fig:h0_nz_supersonic1_soln_ex}
\end{figure}


\section{Nonzero Applied Field, Supersonic Regime}\label{sec:6}
Sectors $\rom{5}$ and $\rom{6}$ satisfy the supersonic condition \eqref{eq:14} in which $v_- < v_+ < 0$ and a contact soliton is developed at the piston boundary. Sector $\rom{7}$ also satisfies the supersonic condition where $0 < v_- < v_+$ and a PDSW emanates from the piston boundary. 

\subsection{Sector $\rom{5}$: $\mathbf{S}^+|\mathbf{RW}$}\label{sec:6a}
The contact soliton is uniquely determined by ~\eqref{eq:19}--~\eqref{eq:20}. With the determined soliton far-field $(n^L,u^L) = (\bar{n},\bar{u})$, the modulation solution is an expansive RW, satisfying \eqref{eq:7} and $r_+ = r_+(\xi)$ where $v_+(r_-,r_+) = \xi = x/\left( t - \bar{t} \right)$. The admissibility condition \eqref{eq:9} is satisfied. Similar to the zero field supersonic solution, no vacuum point is attained. A representative solution is shown in Fig.~\ref{fig:h0_nz_supersonic1_soln_ex}(a) with quantitative agreement between the theoretical prediction and the numerical simulation.

\subsection{Sector $\rom{6}$: $\mathbf{S}^-|\mathbf{RW}\,\mathbf{CDSW}^+$}
The depression contact soliton is uniquely determined by \eqref{eq:19}--\eqref{eq:20} with far-field $(n^L, u^L) = (\bar{n},\bar{u})$ breaking the convexity condition \eqref{eq:5}, leading to a $\mathrm{RW}\,\mathrm{CDSW}^+$ composite wave satisfying $r_- = r_-^R$, $v_3 = (r_-^R,r_+^R,\lambda_3,\lambda_3) = \xi = x/\left( t - \bar{t}\right)$, and $\lambda_4 = \lambda_3$. A representative solution is shown in Fig.~\ref{fig:h0_nz_supersonic1_soln_ex}(b). Same as in sector $\rom{4}$, the onset of vacuum, when the algebraic soliton of the CDSW portion reaches 1, is independent of $u_0$ and happens at $h_0 = 1/\sqrt{2}$. This thresh The pink-shaded region in Fig.~\ref{fig:phase_diagram} indicates a vacuum state is present in the solution.  A vacuum solution from this sector is shown in Fig.~\ref{fig:h0_nz_supersonic1_soln_ex}(c).

\subsection{Sector VII: PDSW$^+$|DSW$^+$}
\begin{figure}[h!]
\centering
\includegraphics[scale=0.35]{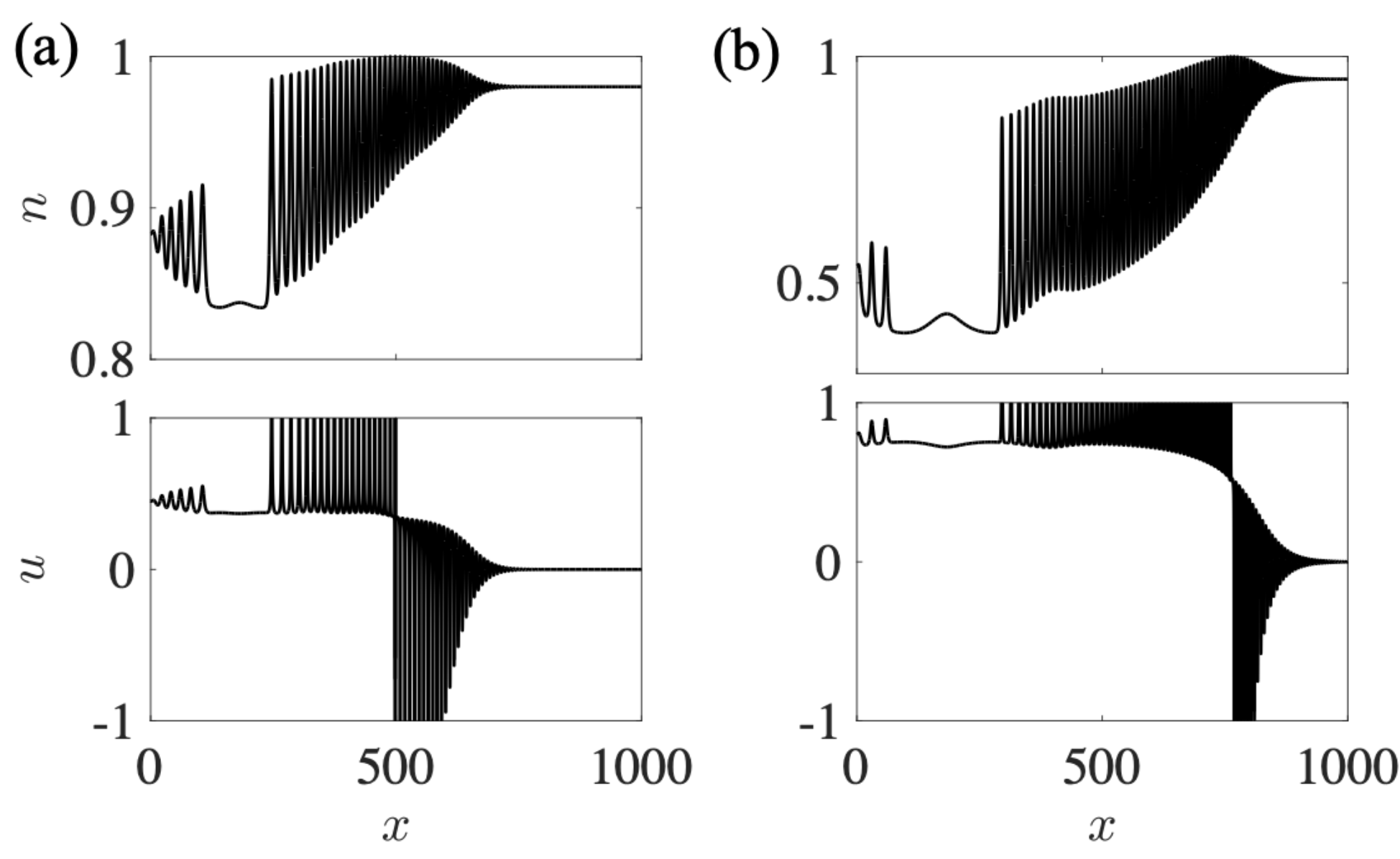}
\caption{(a) $\mathrm{PDSW}|\mathrm{DSW}^+$ solution in sector $\rom{7}$ with $u_0 = 0.45$ and $h_0 = 0.98$. (b) Supercritical solution in sector $\rom{3}$ with $u_0 = 0.8$ and $h_0 = 0.95$.}
\label{fig:h0_nz_supersonic_supercritical_soln_ex}
\end{figure}

The supersonic condition $|u_0| > u_{\mathrm{cr}}(n^L)$ in sector $\rom{7}$ is $0 < v_- < v_+$. This positive velocity configuration is different from all other supersonic sectors with negative dispersionless velocities. It gives rise to a PDSW \cite{marchant1991} at the piston edge. For this sector, we focus on the qualitative identification of the solution features with the support of simulations. As we observed numerically (see Fig.~\ref{fig:h0_nz_supersonic_supercritical_soln_ex}(a) for example), the PDSW is led by a soliton at its right edge and terminates on the left at the piston boundary without reaching the small amplitude limit. The intermediate state connecting the PDSW and a DSW-type wave demonstrates slow oscillations that possibly is not a constant plateau and requires additional analysis. Without the PDSW far-field determined, we are unable to determine the modulation solution. Note that a vacuum point is present inside the solution, consistent with our prediction in Fig.~\ref{fig:phase_diagram}.

We have numerically confirmed that along the sonic curve bounding the subsonic sector $\rom{2}$, where the system remains convex, there is no PDSW emerging from the piston boundary. However, within the nonconvex subsonic sector $\rom{3}$ when near the sonic curve at the sector $\rom{7}$ boundary, we numerically observed that a PDSW develops at the piston boundary as shown in Fig.~\ref{fig:h0_nz_supersonic_supercritical_soln_ex}(b). Consequently, the predicted sonic boundary between sectors $\rom{3}$ and $\rom{3}$ does not precisely explain this phase change. We have not been able to quantitatively identify the threshold for the occurrence of this phase transition using modulation theory. However, all simulations that we have performed in sector $\rom{7}$ exhibit this PDSW structure.

\section{Conclusion}\label{sec:7}
Using the dispersive hydrodynamic framework, we have analytically classified the piston-like dynamics of a dissipationless easy-plane ferromagnetic channel subject to spin injection at one channel boundary. This framework enables the analytical description of noncollinear magnetic textures beyond the small-amplitude, weakly nonlinear regime. 

Two properties of the system is fundamental to our analysis. First, the piston analogy naturally leads to the investigation of magnetic sub- to supersonic conditions, corresponding to distinct piston boundary behavior: either a constant hydrodynamic flow in the subsonic case, or a soliton or a non-stationary partial DSW (PDSW) both in the supersonic case.  We provided quantitative characterization of the solutions using the modulation theory and qualitative identification of the PDSW  solutions, where a sharp threshold for this behavior is yet to be determined.

Second, the modulation equations exhibit nonconvexity where the modulation velocities coalesce. Adopting the method developed in \cite{ivanov2017dsw}, a non-classical dispersive shock wave solution, a contact DSW (CDSW), is predicted when the system exhibits nonconvexity as a single wave or one component of a composite wave. A distinguishing feature is a short, smooth ramp at the algebraic soliton edge of a CDSW where the soliton connects to a dispersionless (non-oscillatory) portion of the solution.

While our analysis was developed for conservative spin dynamics applicable over short enough time scales, it has intriguing implications for longer times wherein magnetic damping leads to relaxation of the dynamics to a steady configuration. First, rarefaction waves expand in time with negligible oscillations. This implies that such a solution
minimizes the excitation of spin waves in the system. On the contrary,
spin shocks exhibit pronounced oscillations that can reflect many
times in the channel before being quenched by magnetic damping. While
this can be seen as a disadvantage, it is also important to note that
the spin waves excited by a spin shock are launched within a specific
spectral band that is determined by the transition between the left
and right states \cite{conforti_resonant_2014}, opening opportunities
for controllable transport of angular momentum by means of pulsed
injection. Second, we find that a stationary soliton established in
the conservative regime can remain after stabilization via damping,
resulting in the contact soliton-dissipative exchange flow~\cite{iacocca2019def}. Third,
numerical simulations in \cite{hu2021} show that it is also possible
to excite propagating soliton trains that persist, oscillating back and forth in the channel, even in the
presence of damping.  In additional simulataions, we observe here that such solitons are excited
precisely when the originating spin shock contains a CDSW.  These are examples of situations where the transient dynamics impact the transport characteristics of the dissipative exchange flow in equilibrium.

The dispersive hydrodynamic interpretation of ferromagnetic dynamics allows one to adopt a large pool of analytical tools that are traditionally used for classical fluids, which provides new perspectives on the study and understanding of spin dynamics. The dynamical problem studied here has a problem setup that is designed to be experimentally accessible and we expect our methodology to aid the experimental realization of superfluid-like spin transport in the form of nonuniform magnetic textures.

\section{Acknowledgment}\label{sec:8}
M. Hu thanks T. Congy for the helpful discussions. All authors acknowledge support from the U.S. Department of Energy, Office of Science, grant DE-SC0018237. M. Hoefer acknowledges partial support from National Science Foundation DMS-1816934 and M. Hu acknowledges support from the National Institute of Standards and Technology Professional Research Experience Program.

\appendix*
\section{Appendix: Determination of the Physical Wave Pattern Given Riemann Invariants}\label{sec:AA}
\renewcommand{\thefigure}{A\arabic{figure}}
\setcounter{figure}{0}
In this appendix, we present additional information on the characterization of periodic traveling wave solutions to the LL equation \eqref{eq:LL}. 
The LL-Whitham equations in terms of the Riemann invariants $\mathbf{\lambda} = (\lambda_1,\lambda_2,\lambda_3,\lambda_4)$ have been given in \eqref{eq:whitham_eqn}.
The family of traveling waves dynamics satisfy Eq.~\eqref{eq:potential_eqn}. The quartic polynomial $R(n)$ can be written in terms of four roots $\{n_i\}_{i=1}^4$. It can also be expressed in terms of the Riemann invariants $\boldsymbol \lambda$ \cite{ivanov2017dsw}
where
\begin{align}
\begin{split}
R(n) = &n^4 + \f{s_1 + s_3}{f_1}n^3 + s_2 n^2 + (f_1 s_1 - \f{s_1+s_3}{f_1})n\\
           & + \f{1}{4}(s^2_1 - 4 - 4s_2 + 4 f_1^2), \label{eq:R(n)_ap}
\end{split}
\end{align}
\begin{align}
\begin{split}
& s_1 = \displaystyle \sum_i^4 \lambda_i,\  s_2 = \displaystyle \sum_{i<j}^4 \lambda_i \lambda_j,\  s_3 = \displaystyle \sum_{i<j<k}^4 \lambda_i \lambda_j \lambda_k,\\
& s_4 = \lambda_1 \lambda_2 \lambda_3 \lambda_4,\label{eq:si}
\end{split}
\end{align}
\be
\lambda_i' = \sqrt{1- \lambda_i^2}, \ s_4' = \Pi_i^4 \lambda_i'.
\ee
For a given set of Riemann invariants $\boldsymbol \lambda$, there are four possible physical wave patterns corresponding to four possible choices of $f_1$:
\begin{subequations}
\begin{align}
  f_1 &= \pm \sqrt{(1+s_2+s_4+s_4')/2},  \label{eq:f11} \\
  \mathrm{or\ } f_1 &= \pm \text{sgn}(s_1 + s_3)\sqrt{(1+s_2 + s_4 - s_4')/2}, \label{eq:f12}
\end{align}
\end{subequations}
This 4-valued mapping of Riemann invariants to traveling wave profiles implies that the LL-Whitham modulation system is nonconvex. Later, we denote $f_{1a}$ as $f_1$ taking the positive expression in \eqref{eq:f11} and $f_{1b}$ as $f_1$ taking the positive expression in \eqref{eq:f12}. The fluid velocity can be computed from the density as
\be
u(\xi) = -\f{f_1 + \f{s_1}{2}n}{1-n^2}. \label{eq:u_in_n_ap}
\ee


The multi-valued mapping from the Riemann invariants to the roots of the potential function are \cite{ivanov2017dsw}:
\be
\begin{aligned}
n_1 &= -\f{1}{2 f_1} \f{(\lambda_3 - \lambda_2)\tilde{s}_1 + (\lambda_3 - \lambda_1)\tilde{s}_2 - (\lambda_2 - \lambda_1)\tilde{s}_3}
{(\lambda_3 - \lambda_2)\lambda_1' + (\lambda_3 - \lambda_1)\lambda_2 ' - (\lambda_2 - \lambda_1)\lambda_3'},\\
n_2 &= -\f{1}{2 f_1} \f{(\lambda_3 - \lambda_2)\tilde{s}_1 + (\lambda_3 - \lambda_1)\tilde{s}_2 + (\lambda_2 - \lambda_1)\tilde{s}_3}
{(\lambda_3 - \lambda_2)\lambda_1' + (\lambda_3 - \lambda_1)\lambda_2 ' + (\lambda_2 - \lambda_1)\lambda_3'},\\
n_3 &= -\f{1}{2 f_1} \f{(\lambda_3 - \lambda_2)\tilde{s}_1 - (\lambda_3 - \lambda_1)\tilde{s}_2 - (\lambda_2 - \lambda_1)\tilde{s}_3}
{(\lambda_3 - \lambda_2)\lambda_1' - (\lambda_3 - \lambda_1)\lambda_2 ' - (\lambda_2 - \lambda_1)\lambda_3'},\\
n_4 &= -\f{1}{2 f_1} \f{(\lambda_3 - \lambda_2)\tilde{s}_1 - (\lambda_3 - \lambda_1)\tilde{s}_2 + (\lambda_2 - \lambda_1)\tilde{s}_3}
{(\lambda_3 - \lambda_2)\lambda_1' - (\lambda_3 - \lambda_1)\lambda_2 ' + (\lambda_2 - \lambda_1)\lambda_3'},  \label{eq:ni_lambdai}
\end{aligned}
\ee
where $\tilde{s}_i = (s_1 -\lambda_i)\lambda_i ' + s_4\f{\lambda_i '}{\lambda_i} \mp s_4 '\f{\lambda_i}{\lambda_i '}$. The upper sign in $\tilde{s}_i$ is for $f_{1a}$ given by \eqref{eq:f11} and the lower sign is for $f_{1b}$ given by \eqref{eq:f12}. The other two cases when $f_1 < 0$ leads to reordering of the expressions of $n_i$'s, which is $n_i \leftarrow n_{5-i}$, $i = 1,2,3,4$.

\begin{figure}[h!]
\centering
\includegraphics[scale=0.22]{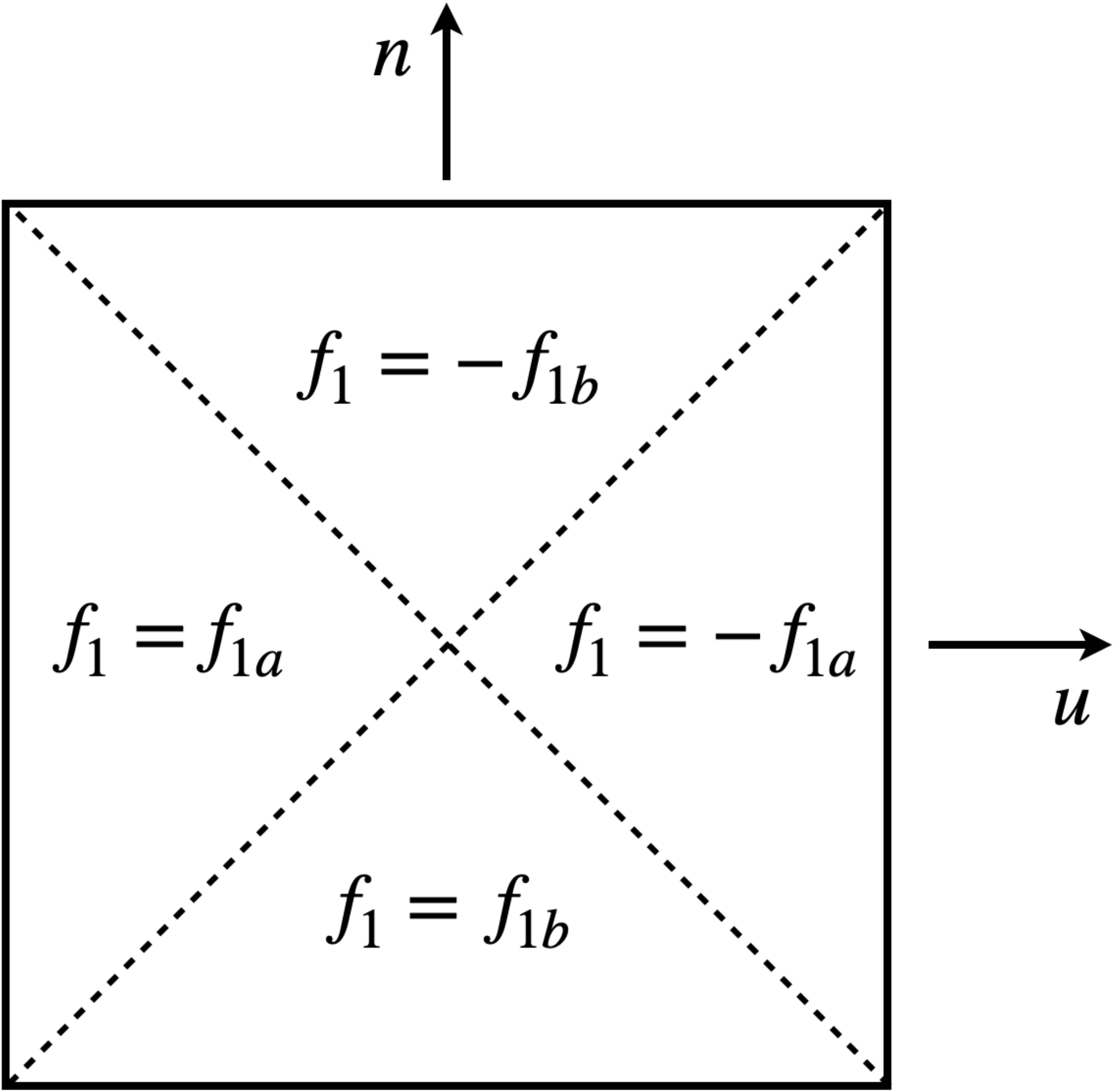}
\caption{The choice of $f_1$ depending on the location of $(u^L,n^L)$.}
\label{fig:f1}
\end{figure}

Depending on which triangle, divided by the diagonal and anti-diagonal of the square $[-1,1] \times [-1,1]$ the left constant state $(u^L,n^L)$ lies in, the choice of $f_1$ is shown in Fig.~\ref{fig:f1}.

\nocite{*}
 \bibliography{LL_piston_arxiv}
\end{document}